\documentclass[prl,floatfix,twocolumn,showpacs,amsmath,amssymb]{revtex4-1}
\usepackage{graphicx} 
\usepackage[dvipdfm]{color}
\usepackage{amsmath}
\usepackage{amsfonts}
\usepackage{amssymb}
\usepackage{wrapfig}

\usepackage{bm}
\usepackage{txfonts}
\newcommand{\subm}[1]{_{\mathrm {#1}}}
\newcommand{\sps}[1]{$^{\mathrm {#1}}$}
\newcommand{\spsm}[1]{^{\mathrm {#1}}}
\newcommand{\etal}{\textit{et~al.}}

\renewcommand{\deg}{^{\circ}}
\newcommand{\Tc}{T\subm{c}}

\newcommand{\Hc}{H\subm{c}}
\newcommand{\Hcc}{H\subm{c2}}

\newcommand{\chin}{\chi\subm{n}}

\newcommand{\chis}{\chi\subm{sc}}

\newcommand{\sro}{Sr$_2$RuO$_4$}

\newcommand{\Hp}{H\subm{P}}

\newcommand{\Hpt}{\tilde{H}_{\mathrm{P}}}

\newcommand{\dSdH}{(\partial S/\partial H)_T}
\newcommand{\Econd}{E_{\mathrm{cond}}}

\newcommand{\Dt}{\varDelta t}
\newcommand{\Dtu}{\varDelta t_{\uparrow}}
\newcommand{\Dtd}{\varDelta t_{\downarrow}}
\newcommand{\Ts}{T}
\newcommand{\Tsz}{T_0}
\newcommand{\Tb}{T\subm{bg}}
\newcommand{\Tbz}{T\subm{bg0}}
\newcommand{\Tf}{T_{\mathrm{FOT}}}
\newcommand{\Hu}{H_{\mathrm{c2\uparrow}}}
\newcommand{\Hd}{H_{\mathrm{c2\downarrow}}}
\newcommand{\tilt}{\theta}
\newcommand{\uu}{\mid\uparrow\uparrow\rangle}
\newcommand{\dd}{\mid\downarrow\downarrow\rangle}

\newcommand{\eq}[1]{Eq.~\eqref{#1}}
\newcommand{\dQloss}{\mathrm{d'\!}Q\subm{loss}}
\newcommand{\Dloss}{D\subm{loss}}
\newcommand{\Tr}{T\spsm{read}}
\newcommand{\Tsr}{\Ts\spsm{read}}
\newcommand{\SMCE}{S_{\!\mathrm{MCE}}}
\newcommand{\SM}{S_{\!\!M}}
\newcommand{\SC}{S_{\!C}}

\begin{document}

\title{First-Order Superconducting Transition of Sr$_{\bm{2}}$RuO$_{\bm{4}}$}

\author{Shingo~Yonezawa}
\affiliation{Department of Physics, Graduate School of Science, 
Kyoto University, Kyoto 606-8502, Japan}
\author{Tomohiro~Kajikawa}
\affiliation{Department of Physics, Graduate School of Science, 
Kyoto University, Kyoto 606-8502, Japan}
\author{Yoshiteru~Maeno}
\affiliation{Department of Physics, Graduate School of Science, 
Kyoto University, Kyoto 606-8502, Japan}

\email{yonezawa@scphys.kyoto-u.ac.jp}

\date{\today}

\begin{abstract}
By means of the magnetocaloric effect, we examine the nature of the superconducting-normal (S-N) transition of \sro, a most promising candidate for a spin-triplet superconductor.
We provide thermodynamic evidence that the S-N transition of this oxide is of first order below approximately 0.8~K and only for magnetic field directions very close to the conducting plane, in clear contrast to the ordinary type-II superconductors exhibiting second-order S-N transitions.
The entropy release across the transition at 0.2~K is 10\% of the normal-state entropy.
Our result urges an introduction of a new mechanism to break superconductivity by magnetic field.
\end{abstract}

\maketitle

The order of a phase transition provides one of the most fundamental pieces of information of the long-range ordered state accompanied by the phase transition.
In case of superconductivity, the order of the superconducting-normal (S-N) transition in magnetic fields reflects how the superconductivity interacts with the magnetic field and how it is destabilised.
For example, for a type-I superconductor, the in-field S-N transition is a first-order transition (FOT)~\cite{TinkhamText},
because of an abrupt disappearance of the superconducting (SC) order parameter caused by the excess energy for magnetic-flux exclusion.
For a type-II superconductor, in contrast, the in-field S-N transition is ordinarily a second-order transition (SOT)~\cite{TinkhamText}.
In this case, penetration of quantized vortices with accompanying kinetic energy due to orbital currents leads to a continuous suppression of the SC order parameter up to the upper critical field $\Hcc$.
This type of pair-breaking is called the orbital effect.

A well-known exception for type-II superconductivity is the case where the superconductivity is destroyed by the Zeeman spin splitting~\cite{Clogston1962}.
When the spin susceptibility in the SC state, $\chis$, is lower than that in the normal state, $\chin$, the SC state acquires higher energy $\varDelta E\subm{Z} \sim (1/2)(\chin-\chis)\mu_0H^2$ with respect to the normal state, due to the difference of polarizability of the electron spin.
This destroys superconductivity at the Pauli limiting field $\mu_0\Hp \sim [2\mu_0\Econd/(\chin-\chis)]^{1/2}$, where $\varDelta E\subm{Z}$ reaches the SC condensation energy $\Econd$.
Such a pair-breaking effect is called the Pauli effect.
It is theoretically predicted that a strong Pauli effect leads to a first-order S-N transition at temperatures sufficiently lower than the critical temperature $\Tc$~\cite{Matsuda2007JPhysSocJpnReview}.
This prediction has been confirmed in a few spin-singlet superconductors~\cite{Bianchi2002,Radovan2003.Nature.425.51,Lortz2007PhysRevLett}.

The type-II superconductor \sro\ ($\Tc = 1.5$~K) is one of the most promising candidates for spin-triplet superconductors~\cite{Maeno1994,Mackenzie2003RMP,Maeno2011.JPhysSocJpn.81.011009}.
Due to its unconventional superconducting phenomena originating from the orbital and spin degrees of freedom as well as from non-trivial topological aspect of the SC wave function, this oxide continues to attract substantial attention~\cite{Nelson2004.Science.306.1151,Xia2006.PhysRevLett.97.167002,Kashiwaya2011.PhysRevLett.107.077003,Nakamura2011.PhysRevB.84.060512R,Jang2011.Science.331.186}.
The spin-triplet state has been directly confirmed by extensive spin susceptibility measurements by means of the nuclear magnetic resonance (NMR)  using several atomic sites~\cite{Ishida1998.Nature.396.658,Ishida2001.PhysRevB.63.060507R,Murakawa2004.PhysRevLett.93.167004} and the polarized neutron scattering~\cite{Duffy2000.PhysRevLett.85.5412}:
Both experiments have revealed $\chis = \chin$ in the entire temperature-field region investigated.
This means that $\Hp\propto (\chin-\chis)^{-1/2}$ is infinite and the Pauli effect is irrelevant in this material.

Interestingly, several properties of the S-N transition of \sro\ have not been understood for more than 10 years within the existing scenarios for the spin-triplet pairing.
For example, $\Hcc(T)$ is more suppressed than the expected behavior for the orbital effect, when the field is parallel to the conducting $ab$ plane~\cite{Akima1999.JPhysSocJpn.68.694,Deguchi2002,Kittaka2009.PhysRevB.80.174514}.
In addition, several quantities such as the specific heat $C$~\cite{Deguchi2002}, thermal conductivity $\kappa$~\cite{Deguchi2002}, magnetization $M$~\cite{Tenya2006.JPhysSocJpn.75.023702}, exhibit sudden recovery to the normal-state values near $\Hcc$ for $H \parallel ab$ and at low temperatures.

To resolve the origin of such unusual behavior, we performed measurements of the magnetocaloric effect (MCE) of \sro.
The MCE is a change of the sample temperature $\Ts$ in response to a variation of the external magnetic field $H$;
we measure $\Ts$ while sweeping $H$ at a constant rate.
The thermal equation of the MCE is written as~\cite{Rost2009.Science.325.1360}
\begin{align}
\left(\frac{\partial S}{\partial H}\right)_{\!T} = 
- \frac{C}{\Ts}\left(\frac{\mathrm{d} \Ts}{\mathrm{d} H}\right) 
 -\frac{k(\Ts-T\subm{bath})}{\Ts\dot{H}} - \frac{1}{T}\frac{\dQloss}{\mathrm{d}H},
\label{eq:MCE}
\end{align}
where $S$ is the entropy, $C$ is the heat capacity of the sample, $k$ is the thermal conductance between the sample and thermal bath, $\dot{H}$ is the sweep rate of the magnetic field, $T\subm{bath}$ is the temperature of the thermal bath, and $\dQloss$ is the dissipative loss of the sample.
When $k$ is small so that the second term is negligible, the equation reduces to the relation for the conventional adiabatic MCE.
In the other limit where the thermal coupling between the sample and bath is strong, the first term in turn becomes negligible, leading to the ``strong-coupling limit'' relation~\cite{Rost2009.Science.325.1360}
$\dSdH \simeq -(k\Dt/\dot{H}) - T^{-1}(\dQloss/\mathrm{d}H)$
with $\Dt \equiv (\Ts-T\subm{bath})/\Ts$~\cite{NoteSuppleMat}.
In this limit, the measured $\Dt$ is linearly dependent on $(\partial S/\partial H)_T$. 
Thus, it is expected that $\Ts$ and $\Dt$ exhibit peak-like anomalies at a FOT and step-like anomalies at a SOT.
Because of this qualitative difference, the strong-coupling MCE is suitable to distinguish a FOT and a SOT.
We found that our calorimeter indeed works nearly in this strong-coupling limit,
with the first term in \eq{eq:MCE} amounting to at most 10\% of the second term.
We however didn't neglect the first term in the evaluation of the entropy discussed below.

For the present study, we used single crystals of \sro\ grown by the floating-zone method~\cite{Mao2000.MaterResBull.35.1813}: Sample~\#1 weighing 0.684~mg with $\Tc =1.45$~K and Sample~\#2 weighing 0.184~mg with $\Tc=1.50$~K.
The value of $\Tc$ of Sample~\#2 is equal to the ideal $\Tc$ of \sro\ in the clean limit~\cite{Mackenzie1998.PhysRevLett.80.161},
indicating its extreme cleanness.
The MCE was measured using a hand-made sensitive calorimeter. 
Magnetic field was applied using a vector magnet system~\cite{Deguchi2004RSI}.
Details of the experimental method is described in the Supplemental Material~\cite{NoteSuppleMat}.

\begin{figure}
\begin{center}
\includegraphics[width=8.5cm]{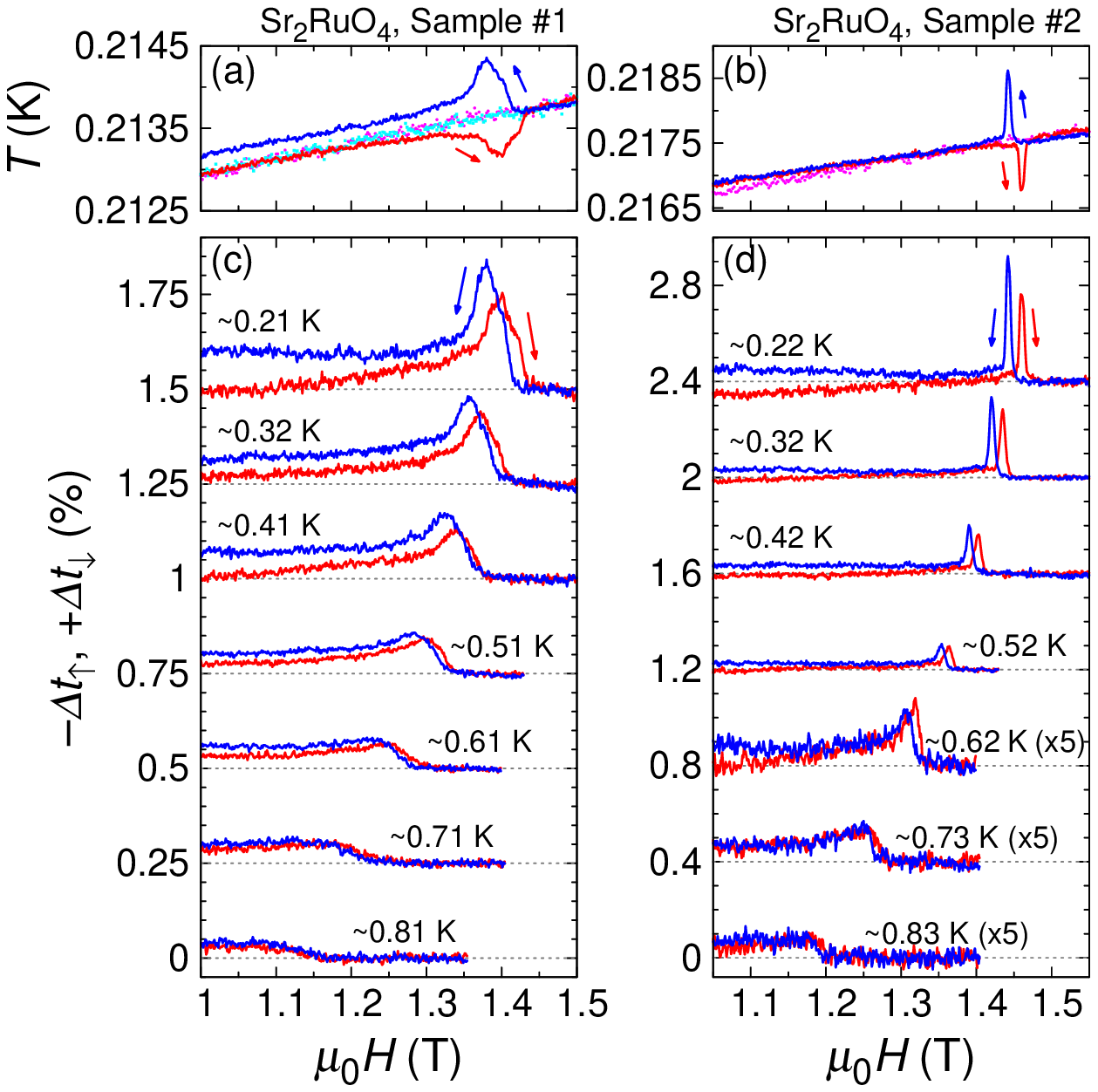}
\end{center}
\caption{
(color online)
(a,b) Representative raw data of the magnetocaloric effect (MCE) of \sro\ for $\Ts\sim 0.2$~K at $H \parallel ab$ ($H\sim [100]$).
The dotted curves indicate $\tilde{t\subm{bg}}\Tsz$ for the up sweep (pink) and the down sweep (cyan), which corresponds to the background contribution~\cite{NoteSuppleMat}.
(c,d) Relative temperature change $-\Dtu$ (red) and $\Dtd$ (blue) due to the MCE for the same field condition at different temperatures.
For clarity, each pair of curves is shifted vertically by 0.25\% and 0.4\% for panels (c) and (d), respectively.
The high-temperature ($T>0.6$~K) data in (d) are multiplied by 5.
\label{fig:T-dependence}
}
\end{figure}

\begin{figure}
\begin{center}
\includegraphics[width=8.5cm]{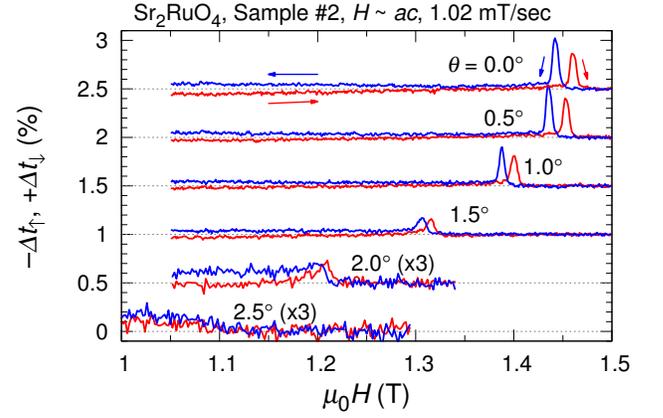}
\end{center}
\caption{
(color online)
Field angle $\tilt$ variation of the magnetocaloric effect of Sample~\#2 at $\Ts\sim 0.2$~K. 
The red and blue curves indicate $-\Dtu(H)$ and $\Dtd(H)$, respectively.
Each curve is shifted vertically by 0.5\% and
the data for $\tilt\geq 2.0\deg$ are multiplied by 3 for clarity.
\label{fig:Theta-dependence}
}
\end{figure}

We first present the MCE for $H \parallel ab$ ($H\sim [100]$) and $\Ts\sim 0.2$ ~K measured at $\mu_0\dot{H}=\pm 1.02$~mT/sec in Figs.~\ref{fig:T-dependence}(a) and (b).
Obviously, $\Ts(H)$ exhibits peak-like behavior near $\Hcc$, rather than a single step-like behavior.
This feature becomes clearer in the background-subtracted $\Dtu(H)$ (up-sweep) and $\Dtd(H)$ (down-sweep) curves shown in Figs.~\ref{fig:T-dependence}(c) and (d)~\cite{NoteSuppleMat}.
The observed peak provides indication of a FOT in \sro.
Note that a slight asymmetry in the MCE signal (i.e. $|\Dtu(H)|<|\Dtd(H)|$) is attributed to the energy dissipation mainly due to vortex motion causing a heating in both the field up-sweep and down-sweep measurements~\cite{NoteAsymmetry}.
More importantly, $\Hcc$ is clearly different between the up-sweep and down-sweep curves.
The difference between the up-sweep onset $\Hu$ and the down-sweep onset $\Hd$ is approximately $ \mu_0\varDelta\Hcc \equiv \mu_0(\Hu-\Hd)=20$~mT for Sample~\#1 and 15~mT for Sample~\#2.
This difference corresponds to 15--20~sec for $\mu_0\dot{H}=1.02$~mT/sec.
The difference cannot be attributed to an extrinsic delay of the temperature measurement, since the delay time of our apparatus is much shorter than 15--20~sec~\cite{NoteThermalRelaxationTime}.
We have also confirmed that a finite $\varDelta \Hcc$ is observed for lower sweep rates such as $\mu_0\dot{H}=\pm 0.2$~mT/sec.
Therefore, this difference of $\Hcc$ is indeed intrinsic, and provides definitive evidence that the S-N transition is a FOT accompanied by supercooling (or possibly superheating).
Note that the very sharp peak in $\Dt(H)$ at $\Hcc$ for Sample~\#2 demonstrates the cleanness and homogeneity of this sample.

Next, we focus on the variation of the MCE with temperature and field angle.
As represented in Figs.~\ref{fig:T-dependence}(c) and (d),
both the peak in $\Dt(H)$ and the supercooling becomes less pronounced as temperature increases.
Around 0.8~K, these features totally disappear and
the S-N transition becomes a SOT as expected for ordinary type-II superconductors.
In Fig.~\ref{fig:Theta-dependence}, we present several MCE curves for fields tilted away from the $ab$ plane toward the $c$ axis by the amount which we define as $\tilt$.
When the field is tilted only by $\sim 2$ degrees,
the FOT features disappear.

\begin{figure}[tb]
\begin{center}
\includegraphics[width=8.5cm]{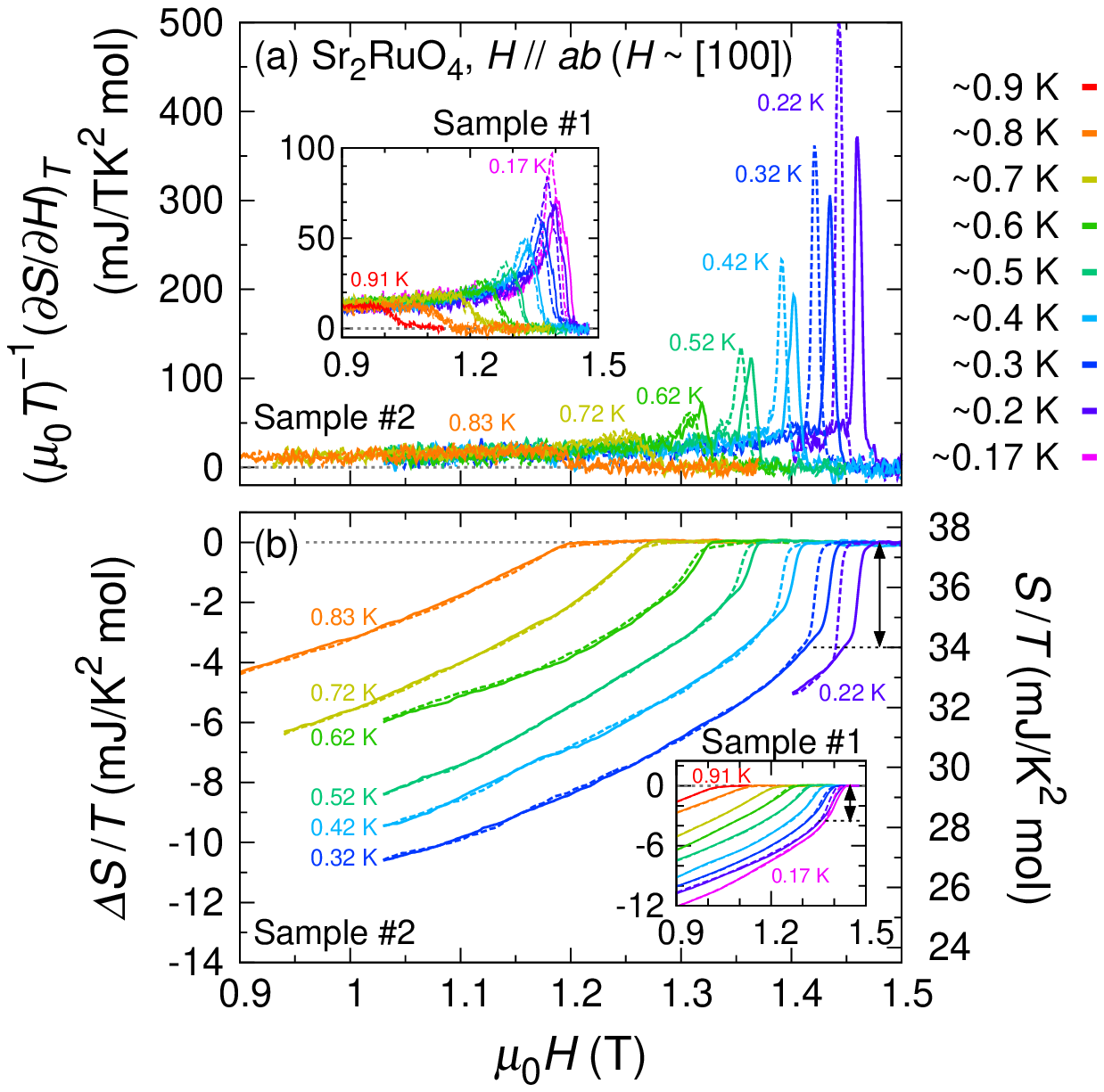}
\end{center}
\caption{
(color online)
(a) Field dependence of $(\mu_0T)^{-1}\dSdH$ of \sro\ deduced from the magnetocaloric effect. 
(b) Field dependence of $\varDelta S/T$.
In both (a) and (b), the main panels present data for Sample~\#2 and the insets for Sample~\#1; the solid and broken curves present up- and down-sweep data, respectively.
The right axis of (b) indicates $S/T$ obtained by assuming $S\subm{n}/T = 37.5$~mJ/K\sps{2}\,mol~\cite{NishiZaki2000JPhysSocJpn}.
The double-headed arrow in (b) illustrates the jump $\delta S/T = -3.5\pm 1$~mJ/K\sps{2}\,mol at the transition.
\label{fig:entropy}
}
\end{figure}

From the MCE data for $H\parallel ab$, we deduce the entropy using \eq{eq:MCE}~\cite{NoteSuppleMat}.
Figure~\ref{fig:entropy}(a) again characterizes the FOT with a huge peak in $\dSdH$ and supercooling/superheating.
In Fig.~\ref{fig:entropy}(b), we present $\varDelta S \equiv S-S\subm{n}= \int^H_{\Hcc} \dSdH\mathrm{d}H$ divided by temperature.
Here, $S\subm{n}$ is the entropy in the normal state.
The total entropy $S$ can be calculated with the assumption $S\subm{n}/T = \gamma\subm{e}$, where $\gamma\subm{e} = 37.5$~mJ/K\sps{2}\,mol is the electronic specific heat coefficient~\cite{NishiZaki2000JPhysSocJpn}.
The jump in $S/T$ across the FOT is approximately $\delta S/T = -3.5\pm 1$~mJ/K\sps{2}\,mol at the lowest measured temperatures. This value of $\delta S/T$ amounts to approximately 10\% of $S\subm{n}/T$, and the latent heat $L=T\delta S$ at 0.2~K is $0.14\pm 0.04$~mJ/mol.
We can check the consistency of this value using the Clausius-Clapeyron equation
$\mu_0\mathrm{d}\Hcc / \mathrm{d}T = -\delta S / \delta M$, where $\delta M$ is the jump in $M$ across the FOT.
Using the values $\mu_0\mathrm{d}\Hcc/\mathrm{d}T \sim -0.20\pm0.05$~T/K estimated from our $\Hu$ data for Sample~\#2 and $\delta M \sim -0.014$~emu/g from the magnetization study~\cite{NoteMagnetization},
we obtain $\delta S/T = -5.2\pm 1.2$~mJ/K\sps{2}\,mol for 0.2~K.
This value reasonably agrees with the value from our MCE experiment.
In addition, $S$ at lower fields also exhibits agreement with other thermodynamic studies~\cite{Deguchi2002,Tenya2006.JPhysSocJpn.75.023702}, as explained in the Supplemental Material~\cite{NoteSuppleMat}.

\begin{figure}[bt]
\begin{center}
\includegraphics[width=8.5cm]{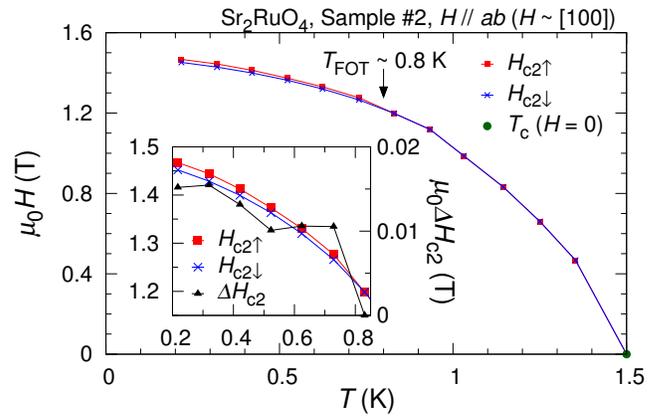}
\end{center}
\caption{
(color online)
Superconducting phase diagram of \sro\ for $H\parallel ab$ ($H\sim [100]$) deduced from the magnetocaloric effect for Sample~\#2. 
The red squares and the blue crosses indicate the onset $\Hcc$ for the up- and down-sweeps, respectively.
The inset presents $\Hcc$ and $\varDelta \Hcc$ (triangles) in the low-temperature region.
\label{fig:phase_diagram}
}
\end{figure}

\begin{figure}[bt]
\begin{center}
\includegraphics[width=8.5cm]{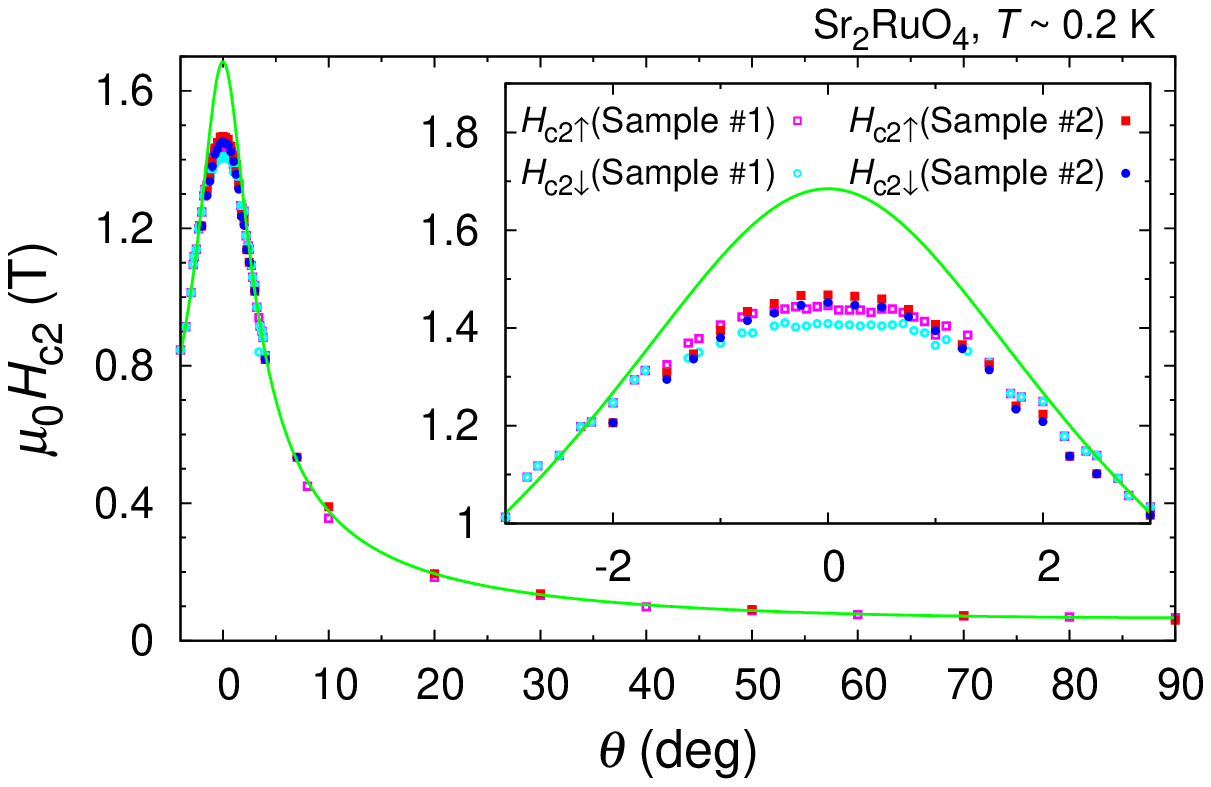}
\end{center}
\caption{
(color online)
Field-angle dependence of $\Hu$ (squares) and $\Hd$ (circles) at $\Ts\sim 0.2$~K.
The green curve indicates the fitting to the data of Sample~\#1 in the range $|\tilt|>2.0\deg$~\cite{Kittaka2009.PhysRevB.80.174514} with the effective mass model $\Hcc(\tilt) = \Hcc(90\deg)/(\sin^2\tilt + \cos^2\tilt/\varGamma^2)^{1/2}$, where $\varGamma$ is the anisotropy parameter and is obtained to be 25 from the fitting.
\label{fig:phase_diagram_angle}
}
\end{figure}

We summarize the present observations in the phase diagrams presented in Figs.~\ref{fig:phase_diagram} and \ref{fig:phase_diagram_angle}.
The region for which the FOT emerges is limited to temperatures below $\Tf\sim 0.8$~K for $\tilt=0\deg$ and field angles within $|\tilt|<2\deg$ for $T\sim 0.2$~K.
Interestingly, the FOT region is included in a wider region in which the behavior of $\Hcc$ cannot be described solely by the conventional orbital effect~\cite{Kittaka2009.PhysRevB.80.174514}: 
$\Hcc(T)$ substantially deviates from the linear behavior and $\Hcc(\tilt)$ cannot be fitted with the effective mass model (Fig.~\ref{fig:phase_diagram_angle}).
These facts indicate that the ordinary orbital effect cannot be a origin of the FOT.

Let us compare the present results with previous observations.
The rapid recoveries of $\kappa/T$~\cite{Deguchi2002}, and $M$~\cite{Tenya2006.JPhysSocJpn.75.023702} near $\Hcc$ for $H\parallel ab$ have been observed in the region where the S-N transition is revealed to be of first order.
Thus, it now turns out that these recoveries are actually consequences of the FOT.
However, supercooling (or superheating) at the S-N transition in \sro\ has never been reported in previous studies. 
This is probably because the supercooled metastable normal state easily nucleates into the SC state. 
Thus, a fast and continuous sweep is helpful to observe the supercooling, rather than point-by-point measurements.
The smallness and cleanness of the present samples have also assisted the observation, because the number of nucleation centers (e.g. surface defects, lattice imperfections) is reduced for small and clean samples.
In contrast to the previous studies on the bulk SC phase, a hysteresis in the in-field S-N transition was observed for the interfacial 3-K phase superconductivity in the \sro-Ru eutectic~\cite{Ando1999.JPhysSocJpn.68.1651,Yaguchi2003.PhysRevB.67.214519}.
Possible relation between this hysteresis and the present observation is worth examining.
We note that the second transition revealed by the $C/T$ measurement~\cite{Deguchi2002} at $H_2$, which is 20--30~mT below $\Hcc$, cannot be attributed to the onset of the FOT. 
Although we have not so far obtained convincing MCE data supporting the $H_2$ anomaly, we need higher experimental resolution to clarify this issue.

In the rest of this Letter, we discuss the origin of the FOT.
As we have explained, the FOT should originate from a pair-breaking effect beyond the conventional orbital effect.
Naively, a possible candidate of such a pair-breaking effect is the Pauli effect~\cite{Machida2008.PhysRevB.77.184515}.
However, in the case of \sro, NMR and neutron studies have revealed $\chis=\chin$~\cite{Ishida1998.Nature.396.658,Duffy2000.PhysRevLett.85.5412,Ishida2001.PhysRevB.63.060507R,Murakawa2004.PhysRevLett.93.167004}.
In particular, the observed negative hyperfine coupling provides strong evidence that NMR correctly detects the spin susceptibility~\cite{Maeno2011.JPhysSocJpn.81.011009}.
In addition, $\chis=\chin$ has been confirmed for different nuclei at several atomic sites.
Other experiments also indirectly support the spin-triplet scenario~\cite{Jang2011.Science.331.186,Nelson2004.Science.306.1151,Xia2006.PhysRevLett.97.167002,Kashiwaya2011.PhysRevLett.107.077003}.
Therefore, the Pauli effect should be absent in \sro\ and the FOT cannot be attributed to the Pauli effect either.  
Thus, an unknown pair-breaking effect, or in other words, a non-trivial interaction between superconductivity and magnetic field, must be taken into account.

For a spin-triplet superconductor with $\chis=\chin$, it is naively expected that the $\dd$ condensate can be mutually converted to the $\uu$ condensate by magnetic field without destroying the SC state.
However, in contrast to the expectation, the present experiment shows that the triplet SC state in \sro\ is no more stable at high fields where the Zeeman splitting is no longer a perturbation.
Indeed, $\Hcc$ for $T\to 0$ nearly matches the field $\Hpt=(2\mu_0\Econd/\chis)^{1/2} \sim 1.4$~T, where {\em the Zeeman spin energy in the SC state} $(1/2)\chis\mu_0H^2$ is equal to $\Econd$~\cite{NoteCondensationEnergy}.
This fact suggests that the Zeeman splitting between $\uu$ and $\dd$ condensates, which has not been considered in the existing theories on the SC phase diagram of \sro~\cite{Agterberg1998.PhysRevLett.80.5184,UdagawaM2005.PhysRevB.71.024511,Kaur2005.PhysRevB.72.144528,Mineev2008.PhysRevB.77.064519,Machida2008.PhysRevB.77.184515}, is a source of the non-trivial coupling between magnetic field and triplet superconductivity.

Let us propose possible mechanisms of the non-trivial interaction.
We can categorise them into microscopic and macro/mesoscopic mechanisms.
The microscopic mechanisms include the pinning of the electron spin direction at certain $k$-points predicted by the band calculation~\cite{Haverkort2008.PhysRevLett.101.026406} and confirmed by the angle-resolved photoemission spectroscopy~\cite{Iwasawa2010.PhysRevLett.105.226406}.
Such a pinning of the spin direction may lead to a constraint in the spin-polarization due to the Zeeman effect.
The closeness of the Fermi energy to the van Hove singularity~\cite{Nomura2000.JPhysSocJpn.69.1856} is also worth considering, because a slight modification of the chemical potential due to the Zeeman effect might disturb the pairing glue.

The macro/mesoscopic mechanisms include possible interactions among the Cooper-pair orbital angular momentum $\bm{L}$, the Cooper-pair spin $\bm{S}$, the vortex vorticity, and the magnetic field.
Indeed, a pair-breaking effect due to $\bm{L}$ was proposed in Ref.~\cite{Lukyanchuk1986.JETPLett.44.233}, although this theory cannot be directly applied as long as the orbital motion is assumed to be purely two dimensional.
As another macro/mesoscopic mechanism, the kinematic polarization discussed in the context of the stability of the half-quantum vortex (HQV) is instructive~\cite{Vakaryuk2009.PhysRevLett.103.057003,Jang2011.Science.331.186}.
It was proposed that a velocity mismatch between $\uu$ and $\dd$ condensates around a HQV results in a shift of the chemical potential of these two condensates due to difference in their kinetic energies and leads to an additional spin polarization coupling to the magnetic field.
By an analogy to this theory, we expect that consideration of kinematics of the condensates in high fields may provide a route to unveil the non-trivial coupling between Cooper-pair and magnetic field.

In summary, our MCE study of \sro\ revealed definitive evidence for a first-order S-N transition in the low-temperature region for fields nearly parallel to the $ab$ plane.
The FOT, not attributable to conventional mechanisms, indicates a non-trivial interaction between spin-triplet superconductivity and magnetic field.
This new information on the bulk superconductivity serves as a basis for investigations of the non-trivial topological nature of the SC wavefunction associated with the ``Majorana-like'' edge modes.
We also anticipate that the abrupt growth of the order parameter across the FOT, accompanied by vortex formation and non-trivial symmetry breaking, should provide a new playground for investigation of novel vortex dynamics, which might be related to quantum turbulence and/or to the Kibble-Zurek mechanism.

We acknowledge T.~Nakamura for his supports; and K.~Ishida, K.~Machida, M.~Sigrist, Y.~Yanase, D.~F.~Agterberg, T.~Nomura, K.~Tenya, K.~Deguchi for useful discussions.
We also acknowledge KOA Corporation for providing us with their products for the calorimeter.
This work is supported by a Grant-in-Aid for the Global COE ``The Next Generation of Physics, Spun from Universality and Emergence'' and by Grants-in-Aids for Scientific Research (KAKENHI 22103002, 23540407, and 23110715) from MEXT and JSPS.

\clearpage

\renewcommand{\theequation}{S\arabic{equation}}
\setcounter{equation}{0}
\renewcommand{\thefigure}{S\arabic{figure}}
\setcounter{figure}{0}
\renewcommand{\thetable}{S\arabic{table}}
\setcounter{table}{0}
\makeatletter
\c@secnumdepth = 2
\makeatother

\onecolumngrid

\begin{center}
{\large \textmd{Supplemental Material for}\\[0.3em]
{\bfseries First-Order Superconducting Transition of Sr$_{\bm{2}}$RuO$_{\bm{4}}$}}
\end{center}

\vspace{0.3em}

\begin{center}
{\large Shingo~Yonezawa, Tomohiro~Kajikawa, Yoshiteru~Maeno\\[0.2em]}
Department of Physics, Graduate School of Science, 
Kyoto University, Kyoto 606-8502, Japan
\end{center}

\email{yonezawa@scphys.kyoto-u.ac.jp}

\section{Difference between the adiabatic and strong-coupling magnetocaloric effects}

\begin{wrapfigure}[15]{r}[0pt]{80mm}
\begin{center}
\includegraphics[width=7cm]{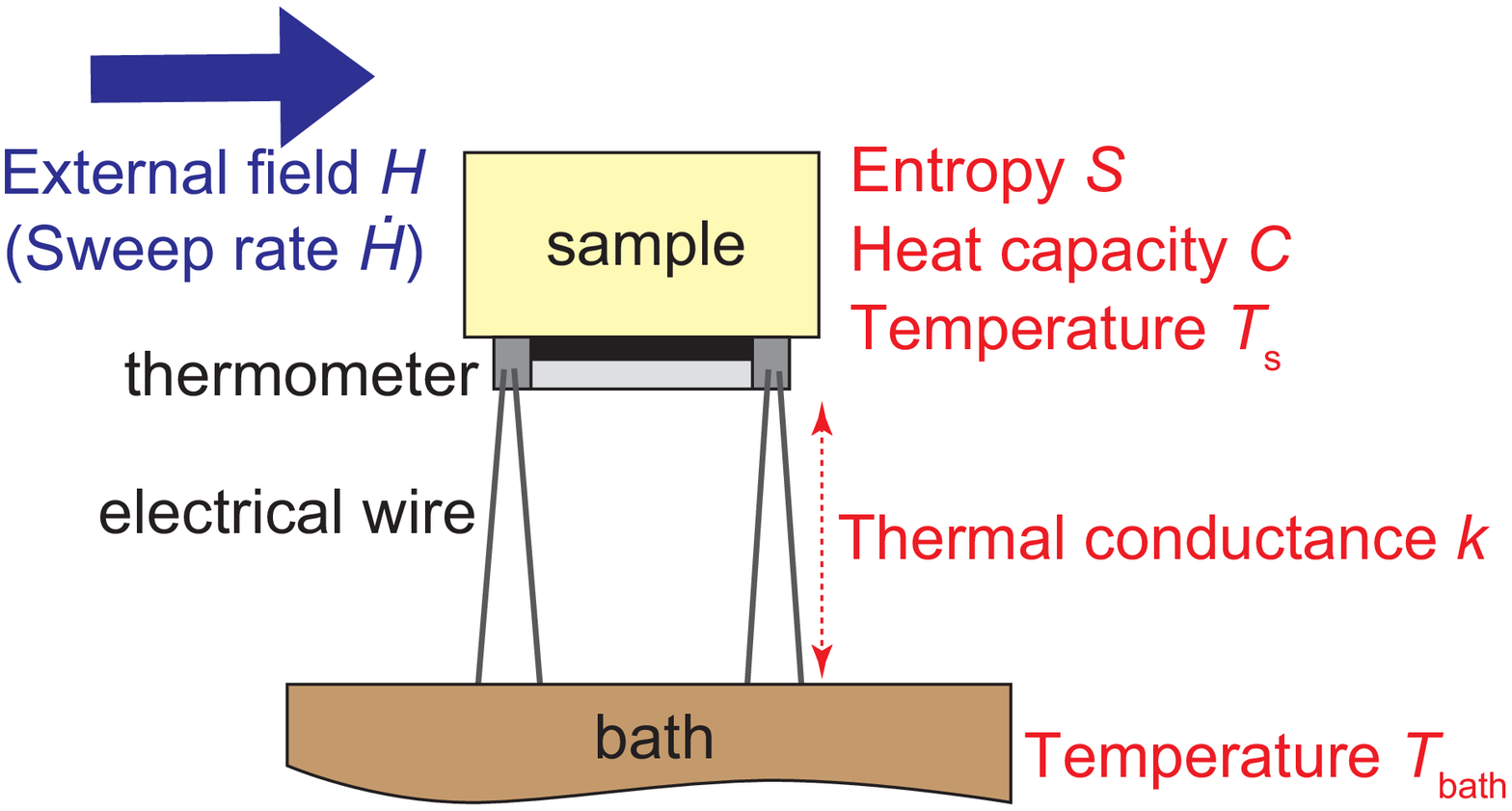}
\caption{
Schematic drawing of a typical setting for magnetocaloric effect measurements.
\label{fig:mag-cal-schematics}
}
\end{center}
\end{wrapfigure}


As already mentioned in the main text, the basic formula for the magnetocaloric effect (MCE) is expressed as:
\begin{align}
\left(\frac{\partial S}{\partial H}\right)_{\!T} = 
- \frac{C}{\Ts}\left(\frac{\mathrm{d} \Ts}{\mathrm{d} H}\right) 
 -\frac{k(\Ts-T\subm{bath})}{\Ts\dot{H}} - \frac{1}{\Ts}\frac{\dQloss}{\mathrm{d}H},
\label{eq:MCE.SM}
\end{align}
where $S$ is the entropy, $C$ is the heat capacity of the sample, $k$ is the thermal conductance between the sample and thermal bath, $\dot{H}$ is the sweep rate of the magnetic field, $\Ts$ is the sample temperature, $T\subm{bath}$ is the temperature of the thermal bath, and $\dQloss$ is the dissipative loss of the sample (see Fig.~\ref{fig:mag-cal-schematics}).
We call the case where the second term is negligibly small as the adiabatic MCE (AMCE) and where the first term is negligibly small as the strong-coupling MCE (SMCE).

\begin{figure}[htb]
\begin{center}
\includegraphics[width=15cm]{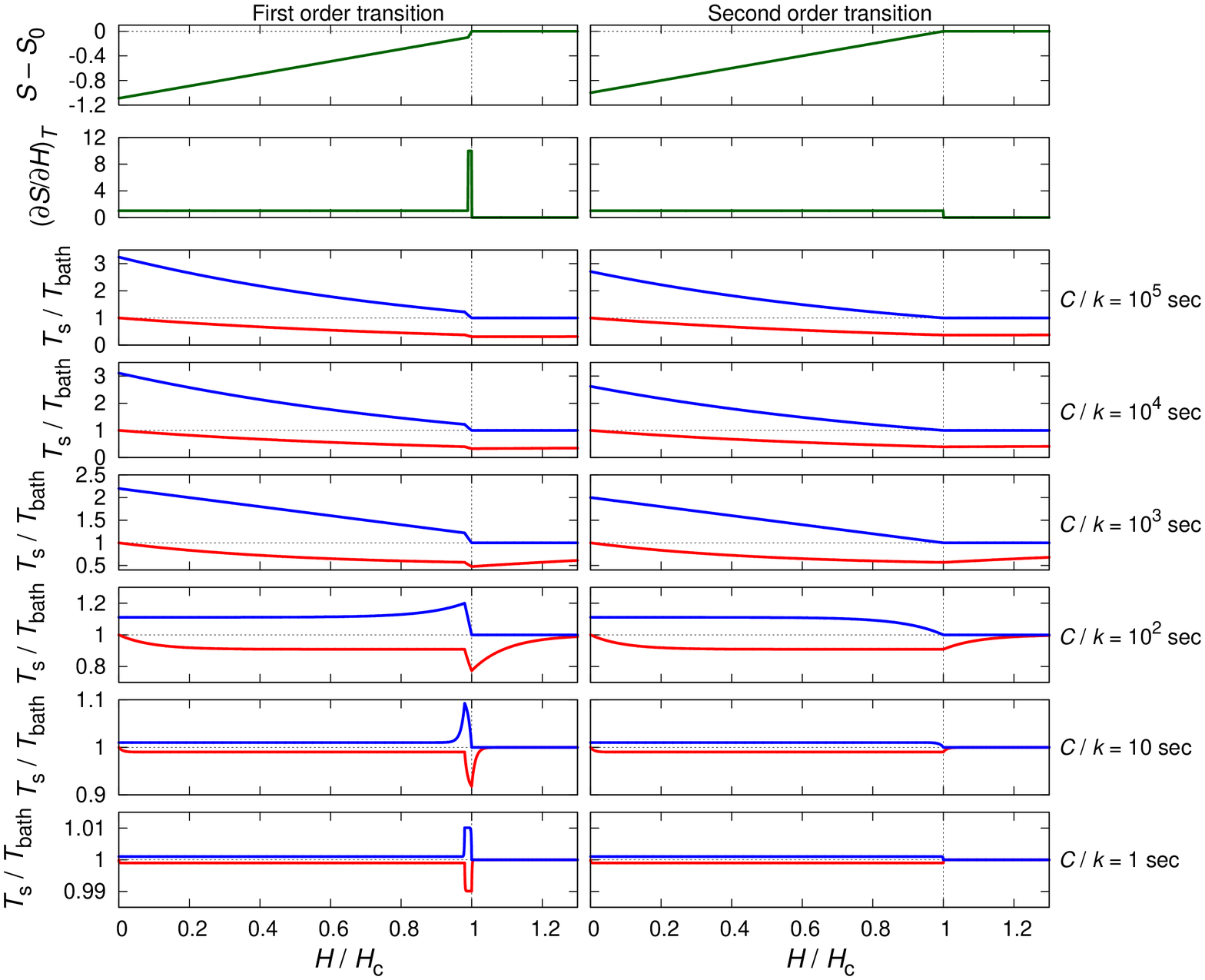}
\end{center}
\caption{
Demonstration of differences between the adiabatic magnetocaloric effect (AMCE) and the strong-coupling magnetocaloric effect (SMCE) for a first-order transition (left column) and a second-order transition (right column).
The assumed functional form of $S(H)$, as well as its field derivative $\dSdH$, is presented in the top panels. 
The calculated MCE responses with different values of the sample-bath relaxation time $C/k$ are presented in the rest of the panels. 
The red and blue curves indicate field-up sweeps and field-down sweeps, respectively. 
The system approaches the AMCE limit for larger $C/k$, whereas it approaches the SMCE limit for smaller $C/k$.
The parameters $\dot{H}$ and $C$ are chosen to be $\dot{H}/H\subm{c}=0.001$~sec\sps{-1} and $CT\subm{bath}/(AH\subm{c})=1.0$.
Note that, for $C/k=10^3$~sec, a deviation from the adiabatic limit shows up, for example, in the field dependence of $\Ts$ for $H>H\subm{c}$.
\label{fig:MCE-simulation}
}
\end{figure}

To demonstrate the difference in the behavior of $\Ts(H)$ between the AMCE and SMCE, we solve \eq{eq:MCE.SM} numerically by assuming functional forms of $S(H)$, $C(H)$, $k(H)$ and $\dQloss(H)$. 
For simplicity, we here only treat the case $k(H) = \text{const.}$ and $\dQloss(H)=0$.
We further assume $C(H)=\text{const.}$ in order to extract the essential difference between the AMCE and SMCE, although $C$ in reality should depends on $H$ as well as on $\Ts$, especially near a phase transition.
For $S(H)$, we adopt simple linear field dependences to simulate a first-order transition (FOT) and a second-order transition (SOT) as shown in the top four panels of Fig.~\ref{fig:MCE-simulation}:
\begin{align}
S\subm{FOT}(H) = 
\begin{cases}
\,A(H-H\subm{c*})+B(H\subm{c*}-H\subm{c})+S_0,\,\,\,\,\,&(H < H\subm{c*})\\
\,B(H-H\subm{c})+S_0,&(H\subm{c*} < H < H\subm{c})\\
\,S_0,               &(H\subm{c} < H);
\end{cases}
\end{align}
and
\begin{align}
S\subm{SOT}(H) = 
\begin{cases}
\,A(H-H\subm{c})+S_0,\,\,\,\,\,&(H < H\subm{c})\\
\,S_0,                         &(H > H\subm{c}).
\end{cases}
\end{align}
In these equations, $H\subm{c}$ denotes the critical field of the phase transition and $H\subm{c*}$ is the onset field of the FOT introduced to simulate a realistic imperfection with a finite broadening of the FOT. 
Here, we used the FOT width $H\subm{c}-H\subm{c*}=0.01H\subm{c}$. 
The coefficients $A$ and $B$ are the slopes of $S(H)$ in the low-field phase and within the FOT region, respectively. 
We here assume $B/A = 10$, i.e. a 10-times steeper slope within the FOT region than in the low-field phase. 
The entropy in the high-field phase is assumed to be a constant $S_0$.

The results of the calculation for different values of $k$ are presented in Fig.~\ref{fig:MCE-simulation}. 
As the sample-bath relaxation time $C/k$ becomes smaller, the system evolves from the AMCE to SMCE. 
As one can see, $\Ts(H)$ is almost proportional to $\dSdH$ for the SMCE as expected from \eq{eq:MCE.SM}. 
Thus, the qualitative difference between a FOT (peak in $\Ts(H)$) and a SOT (step in $\Ts(H)$) is quite clear,
although the MCE signal $\Ts(H)/T\subm{bath}$ is relatively small.
In contrast, for the AMCE, qualitative difference between the shapes of a FOT curve and of a SOT curve becomes less pronounced, 
whereas the available MCE signal can be very large compared to that of the SMCE.

To conclude this section, we demonstrate that the SMCE is a reliable method to distinguish a FOT from a SOT in spite of relatively small changes in the sample temperature.
For the SMCE, such a determination can be done just from the qualitative shape of the raw $\Ts(H)$ curve even without detailed analyses of the curve.

\section{Details of the experimental method}

For the present study, we used single crystals of \sro\ grown by the floating-zone method~\cite{Mao2000.MaterResBull.35.1813.SM}: Sample~\#1 weighing 0.684~mg with $\Tc =1.45$~K and Sample~\#2 weighing 0.184~mg with $\Tc=1.50$~K.
The value of $\Tc$ of Sample~\#2 is equal to the ideal $\Tc$ of \sro\ in the clean limit~\cite{Mackenzie1998.PhysRevLett.80.161.SM},
indicating its extreme cleanness.
The samples were cut, cleaved, and polished; their $\Tc$ was checked by heat capacity and AC susceptibility measurements.
We found that Sample~\#1 have mosaic structure with a $c$-axis tilting of $\sim0.2\deg$, while Sample~\#2 is free from such mosaicity.
The mosaicity in Sample~\#1 does not affect our conclusion.

We developed a sensitive calorimeter for MCE measurements consisting of a small thermometer and a heater made of commercial thick-film RuO$_2$ resistors fixed with low thermal conductance Pt-W wires.
To avoid complication due to the SC transition of solder, we removed the solder coating of the resistors and used silver paste to attach the wires to the resistors.
The calorimeter was cooled with a dilution refrigerator.
We measured $k$ and $C$ separately using the relaxation and ac methods. 
Magnetic field was applied using a vector magnet system consisting of two orthogonal magnets and a horizontal rotating stage~\cite{Deguchi2004RSI.SM}.
For the MCE measurement, we mounted the crystal with its $c$-axis nearly horizontal.
We used the horizontal magnet only because the system does not allow simultaneous field change of the two magnets.
Thus, the accuracy of the field alignment is better than $0.1\deg$ with respect to the $ab$ plane but is $\sim 5$--$10\deg$ for the azimuthal direction within the $ab$ plane.
In order to improve the signal-to-noise ratio of the MCE curves, we repeated field sweeps for 20-30 times and take averages among up and down sweep curves separately.

\section{Process of the evaluation of the entropy}

In this section, we describe details of the process to evaluate the entropy $S(H)$ from the MCE data shown in Fig.~1 of the main paper.

Equation \eqref{eq:MCE.SM} can be rewritten as 
\begin{align}
\left(\frac{\partial S}{\partial H}\right)_T
= -\frac{C}{1-\Dt}\left(\frac{\mathrm{d}\Dt}{\mathrm{d}H}\right) - \frac{k\Dt}{\dot{H}}-\Dloss(\Ts,H),
\label{eq:S-with-dT}
\end{align}
where $\Dt\equiv (\Ts-T\subm{bath})/\Ts$ and $\Dloss(\Ts,H)=(1/\Ts)(\dQloss/\mathrm{d}H)$.
We need to evaluate $\Dt(H)$ and $\Dloss$ from the raw data in order to obtain the entropy $S(H)$.

\begin{figure}[htb]
\begin{center}
\includegraphics[width=15cm]{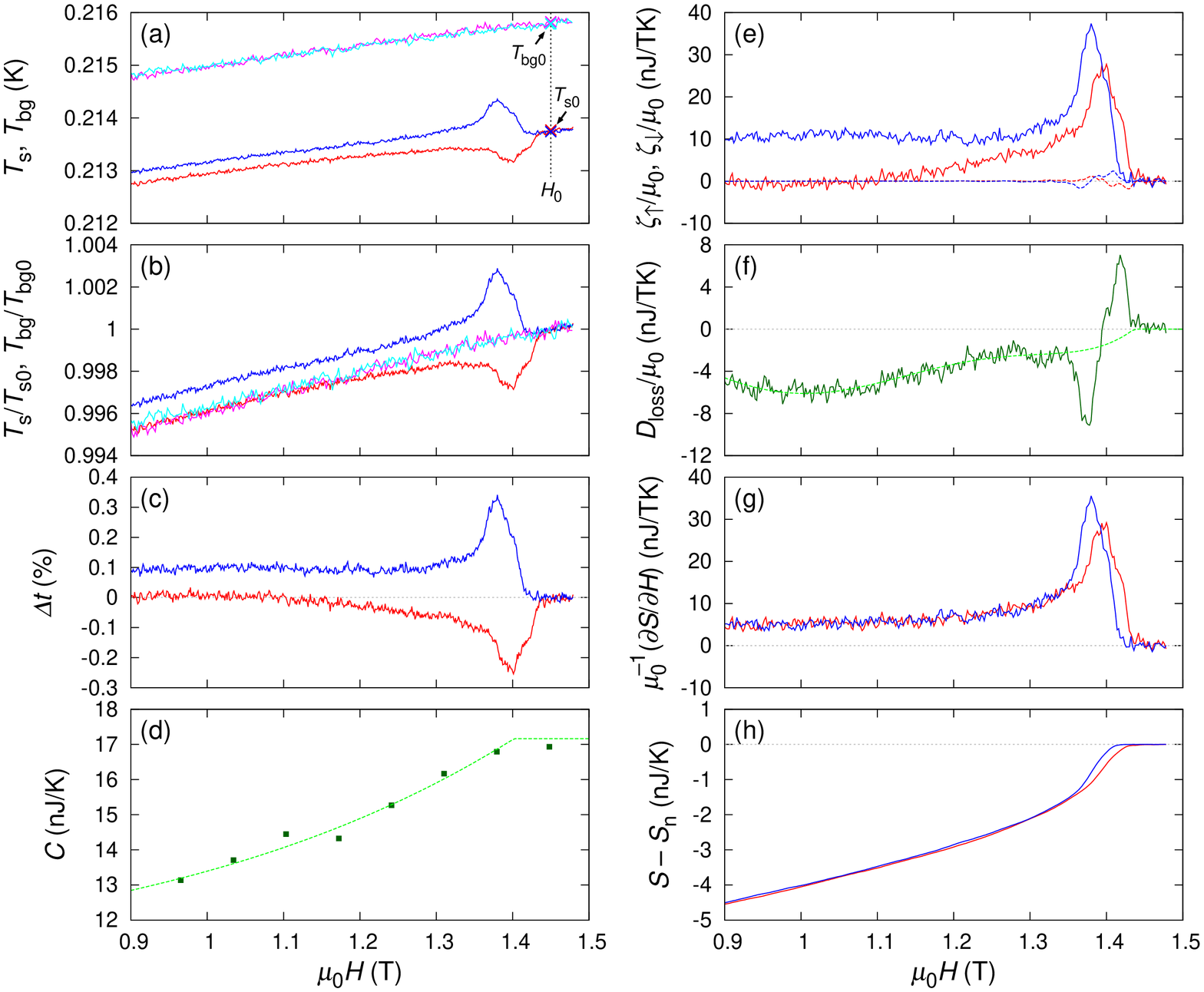}
\end{center}
\caption{
Description of the process of the entropy evaluation at $\Ts\sim 0.21$~K and $\mu_0\dot{H}=1.02$~mT/sec for Sample~\#1. For all data shown here, the red curves present up-sweep data and blue curves present down-sweep data.
(a) Raw data of $\Ts$ and $\Tb$ (up sweep: magenta, down sweep: cyan). $\Tsz=\Ts(1.45~\text{T})$ and $\Tbz=\Tb(1.45~\text{T})$ are indicated with the crosses.
(b) Normalized temperatures $\Ts/\Tsz$ and $\Tb/\Tbz$.
(c) Reduced temperature $\Dt = [(\Ts/\Tsz) - (\Tb/\Tbz)] (\Ts/\Tsz)$.
(d) Heat capacity of the sample plus addenda measured by the relaxation method. The broken curve presents the result of the fitting. This fitted values were used for the entropy-evaluation process.
(e) Nominal entropy derivatives $\zeta_{\uparrow}$ and $\zeta_{\downarrow}$ containing the contribution from the loss term $\Dloss$. The broken curves indicate the contribution of the first term (i.e. the AMCE term) of the right-hand side of \eq{eq:MCE.SM}. The fact that this contribution is at most 10\% of $\zeta_{\uparrow}$ and $\zeta_{\downarrow}$ indicates the present calorimeter nearly works in the strong-coupling limit.
(f) Loss term $\Dloss = (\zeta_{\downarrow} - \zeta_{\uparrow})/2$. The broken curve is the interpolation into the FOT region.
(g) Field derivative of the entropy $\dSdH$.
(h) Entropy evaluated by integrating $\dSdH$.
\label{fig:entropy-evaluation}
}
\end{figure}

In the evaluation process of $\Dt(H)$, we managed to solve an issue on reading errors of thermometers. 
A temperature reading $\Tr$, represented in Fig.~\ref{fig:entropy-evaluation}(a), inevitably contains a certain reading error $\delta T$ with respect to the true temperature $T\spsm{true}$: i.e. $\Tr = T\spsm{true} + \delta T$. 
The error $\delta T$ may originate from errors in the calibration function ($\delta T\spsm{cal}$) and from long-term ($\sim$ days) drifts in the reading in the electronic devices ($\delta T\spsm{drift}$). 
In the present case, since $\delta T$ can be on the same order as the intrinsic temperature variation due to the MCE, we need to eliminate errors originating from $\delta T$ for accurate evaluation of $S(H)$.

The only way to cancel out unknown calibration errors $\delta T\spsm{cal}$ in $\Ts$ and $T\subm{bath}$ is to use a temperature measured with the same thermometer as $\Ts$, instead of $T\subm{bath}$. Thus, we adopt $\Tb$, which is the sample temperature for fields away from the $ab$ plane. 
In the present study, $\Tb$ was measured in fields $20\deg$ away from the $ab$ plane; such a tilt of the magnetic field reduces $\Hcc$ down to 0.2~T at 0.2~K. 
Using $\Tb$, the extrinsic (i.e. background and normal-state) contribution of the MCE signal can be subtracted and we can evaluate the entropy change only due to the superconductivity. 

To minimize contributions from $\delta T\spsm{drift}$, we re-define $\Dt$ using the normalized temperatures $\Ts/\Tsz$ and $\Tb/\Tbz$ (Fig.~\ref{fig:entropy-evaluation}(b)) as
\begin{align}
\Dt(H) \equiv \frac{ \Ts(H)/\Tsz - \Tb(H)/\Tbz }{\Ts(H)/\Tsz},
\label{eq:t-definition}
\end{align}
with $\Tsz\equiv\Ts(H=H_0)$ and $\Tbz\equiv\Tb(H=H_0)$, where $H_0$ is a certain field $H=H_0$ above $\Hcc$ for $H\parallel ab$. 
In the case of Fig.~\ref{fig:entropy-evaluation}, we chose $\mu_0H_0=1.45$~T.
We have indeed confirmed that the reading errors are canceled out and $\Dt \simeq 1 - \Tb\spsm{true}/\Ts\spsm{true}$ up to the first order in small values, as we explain in detail in Appendix~\ref{sec:app:error}.
The obtained $\Dt$ is presented in Fig.~\ref{fig:entropy-evaluation}(c).

Other two important quantities, $C$ and $k$, are determined from separate measurements using the relaxation-time method and the ac method. 
The obtained $C$ data (Fig.~\ref{fig:entropy-evaluation}(d)) is consistent with the previous works~\cite{Deguchi2002.SM,Deguchi2004.PhysRevLett.92.047002.SM,Deguchi2004.JPhysSocJpn.73.1313.SM}. 
We found that $k$ was field-independent within our experimental resolution in the present field range and we adopt $k=11.4$~nW/K for the analysis of the data in Fig.~\ref{fig:entropy-evaluation}.

Next, we need to evaluate the loss term $\Dloss$, which leads to an asymmetric ($|\Dtu|\neq|\Dtd|$) MCE signal.
For the present case, dealing with a type-II superconductor, 
$\Dloss$ originates from heating due to incoming and escaping motion of vortices associated with both increasing and decreasing fields. 
For the present case, the MCE signal is indeed asymmetric (i.e. $|\Dtu|<|\Dtd|$) in the superconducting (SC) state,
even down to fields much smaller than $\Hcc$ (see Fig.~1, and Fig.~\ref{fig:entropy-evaluation}).
Note that, in general, a FOT is often accompanied with an energy dissipation~\cite{Silhanek2006.PhysRevLett.96.136403.SM}.
However, for the present case, 
it is difficult to conclude that the energy dissipation characteristic of the FOT has any substantial contribution to the MCE signal in this case,
because we did not detect additional asymmetry near the FOT as shown in Fig.~\ref{fig:entropy-evaluation}(f)

The loss term $\Dloss$ is estimated by taking an average of the up-sweep and down-sweep results: We first evaluate the ``nominal'' entropy derivatives
\begin{align}
\zeta_{\uparrow} \equiv 
-\frac{C}{1-\Dtu}\left(\frac{\mathrm{d}\Dtu}{\mathrm{d}H}\right) - \frac{k\Dtu}{\dot{H}}
\label{eq:entropy-up}
\end{align}
for up-sweep data ($\dot{H}>0$) and
\begin{align}
\zeta_{\downarrow}
\equiv -\frac{C}{1-\Dtd}\left(\frac{\mathrm{d}\Dtd}{\mathrm{d}H}\right) - \frac{k\Dtd}{\dot{H}}
\label{eq:entropy-down}
\end{align}
for down-sweep data ($\dot{H}<0$), as shown in Fig.~\ref{fig:entropy-evaluation}(e).
Then $\Dloss$ (Fig.~\ref{fig:entropy-evaluation}(f)) is obtained by
\begin{align}
\Dloss(T,H) = \frac{1}{2} \left(\zeta_{\downarrow} - \zeta_{\uparrow}\right).
\label{eq:Dloss}
\end{align}
This process is quite similar to that used in Ref.~\cite{Lortz2007.PhysRevB.75.094503.SM}. 
This is based on the assumption that $\Dloss$ is independent of the sweep direction.
We should be careful, here, that the right-hand side of \eq{eq:Dloss} can be finite even if $\Dloss=0$ near a FOT because of supercooling/superheating. Thus, we need to estimate $\Dloss$ within the FOT regions by an interpolation using a simple polynomial function as indicated by the broken curve Fig.~\ref{fig:entropy-evaluation}(f). This interpolation is performed so that the entropy-conservation law is satisfied between above and far below $\Hcc$.

With the estimated $\Dloss$, $\dSdH$ can be obtained using Eq.~\eqref{eq:S-with-dT} as shown in Fig.~\ref{fig:entropy-evaluation}(g). 
Then, as presented in Fig.~\ref{fig:entropy-evaluation}(h), the entropy $S(H)$ can be calculated by the integration
\begin{align}
S(H)-S\subm{n} = \int^H_{\Hcc} \left(\frac{\partial S}{\partial H}\right)_{T} \mathrm{d}H,
\end{align}
where $S\subm{n}$ is the normal-state entropy.
Note that the evaluations of $\dSdH$ and $S$ become less accurate when $\Dloss$ dominates the MCE.

\section{Comparison of Sample~\#1 and Sample~\#2}

\begin{figure}[htb]
\begin{center}
\includegraphics[width=15cm]{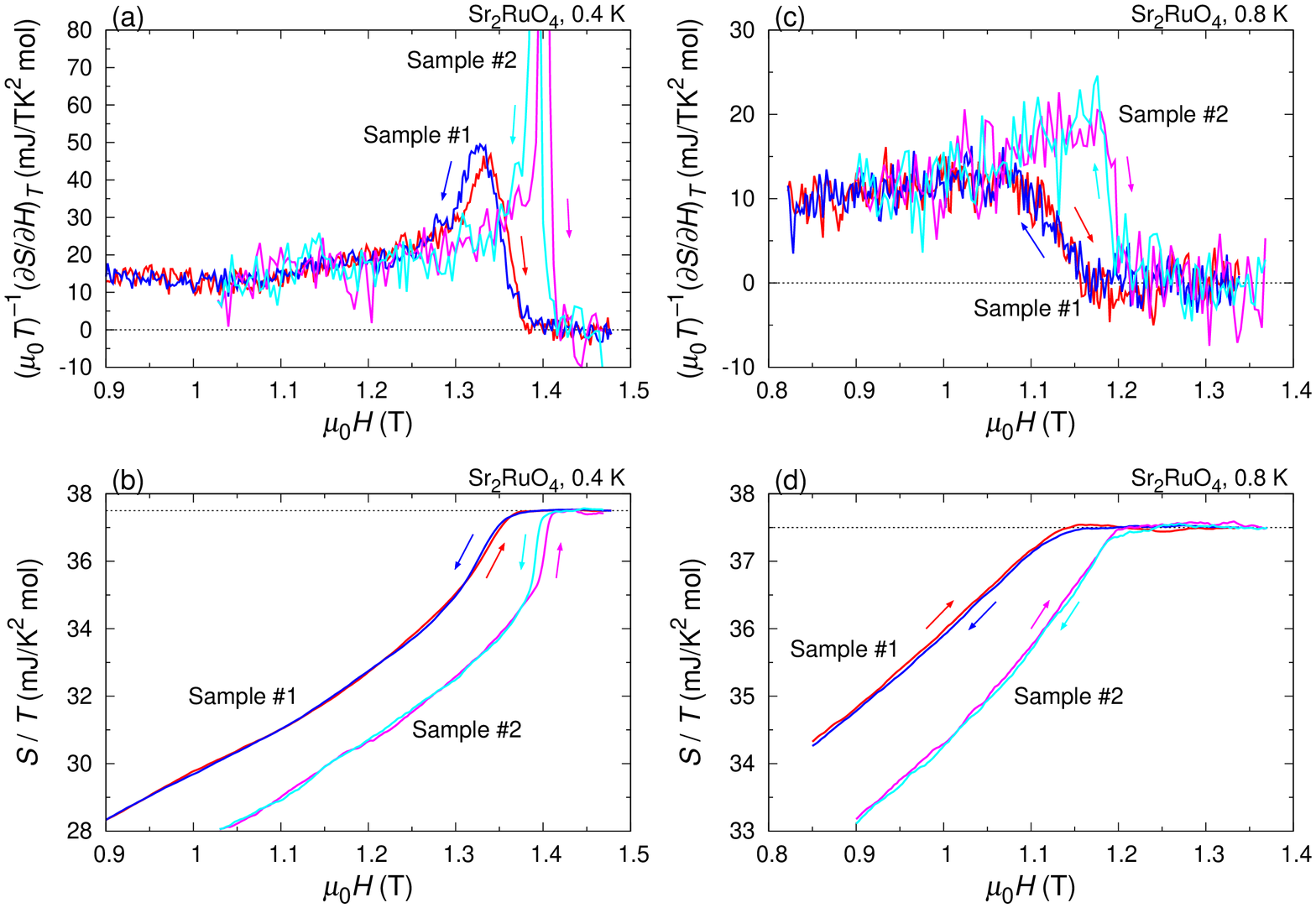}
\end{center}
\caption{
Comparison of the entropy $S(H)$ and its field derivative $\dSdH$ for Sample~\#1 and \#2,
at (a,b) 0.4~K and (c,d) 0.8~K. 
In all panels, the red and blue curves indicate the values for Sample~\#1, and the magenta and cyan curves indicate the values for Sample~\#2.
The $S(H)$ curves in (b) and (d) are obtained with an assumption that the normal state entropy $S\subm{n}$ is 37.5~mJ/K\sps{2}\,mol~\cite{NishiZaki2000JPhysSocJpn.SM} for both samples, as indicated with the dotted lines.
\label{fig:entropy-compare_0.8K}
}
\end{figure}

In order to demonstrate the validity of our entropy evaluation, we here compare results for two \sro\ crystals described in the main text: Sample~\#1 (the SC transition temperature $\Tc=1.45$~K with a broader transition in magnetic fields) and Sample~\#2 ($\Tc=1.50$~K with an extremely sharp transition even in magnetic fields).
When the loss term $\Dloss$ is smaller than the intrinsic contributions, we can accurately evaluate $\dSdH$ as well as $S$ both for Sample~\#1 and Sample~\#2.
As presented in Figs.~\ref{fig:entropy-compare_0.8K}(a) and (c), $\dSdH$ of both samples agrees with each other below 1.25~T at 0.4~K and below 1.05~T at 0.8~K, although deviations originating from the difference in $\Hcc$ and the sharpness of the SC transition have been observed near $\Hcc$. 
Accordingly, the $S(H)$ curves, drawn with an assumption that the normal-state entropy $S\subm{n}/ = 37.5$~mJ/K\sps{2}~\cite{NishiZaki2000JPhysSocJpn.SM} is common between Sample~\#1 and \#2,
reflect the similarity and difference in $\dSdH$.
The data at 0.3~K shown in Fig.~\ref{fig:entropy-compare} also exhibits good agreement considering the characteristics of each sample.
This agreement between different samples indicates the validity of our analyses.

At lower temperatures, the contribution of $\Dloss$ becomes more significant. For Sample~\#2, since $\Dloss$ dominates the MCE signals as shown in Fig.~1 of the Main Text, it is quite difficult to accurately evaluate the intrinsic $\dSdH$ except near $\Hcc$, where the intrinsic contribution is still larger than $\Dloss$.
Indeed, the jump in the entropy across $\Hcc$, $\delta S/T \sim 3.5$~mJ/K$^2$\,mol, is consistent between the two samples, as already mentioned in the Main Text.

\section{Comparison with other thermodynamic studies}

In this section, we compare the entropy obtained from our MCE results, $\SMCE$, with those from other thermodynamic studies. 

\begin{figure}[htb]
\begin{center}
\includegraphics[width=15cm]{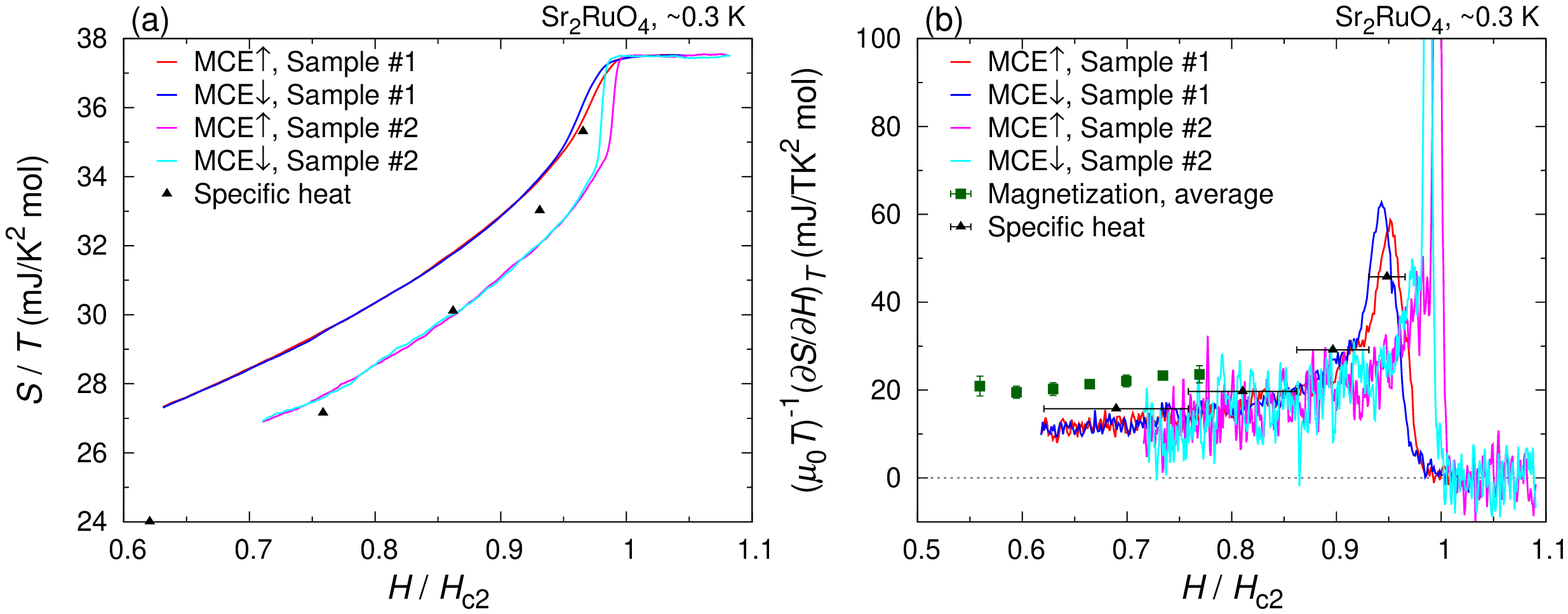}
\end{center}
\caption{
Comparison of the MCE results with previous thermodynamic studies. 
(a) Comparison of $S(H)$ obtained from the MCE and specific heat~\cite{Deguchi2002.SM}. 
(b) Comparison of $\dSdH$ obtained from the MCE, magnetization~\cite{Tenya2006.JPhysSocJpn.75.023702.SM}, and specific heat~\cite{Deguchi2002.SM}.
\label{fig:entropy-compare}
}
\end{figure}

The specific heat $C$ has a relation to the associated entropy $\SC$ as
\begin{align}
\SC = \int^T_0\!\left(\frac{C}{T}\right)\mathrm{d}T.
\end{align}
To evaluate $\SC$, which are plotted in Fig.~\ref{fig:entropy-compare}(a), we first extrapolated the $C(T)/T$ data in Ref.~\cite{Deguchi2002.SM} to 0~K using a polynomial function. We then integrated $C/T$ up to 0.32~K. 
We also plot $(\partial \SC/\partial H)_T$ in Fig.~\ref{fig:entropy-compare}(b) obtained by simply taking two-point slopes of the $\SC(H)$ data, together with $(\partial\SMCE/\partial H)_T$.
As demonstrated in Figs.~\ref{fig:entropy-compare}(a) and (b), $\SC$ and $\SMCE$, as well as $(\partial \SC/\partial H)_T$ and $(\partial\SMCE/\partial H)_T$, reasonably agrees with each other in the present field range.

The Maxwell's relation yields the relation between the magnetization $M$ and the associated entropy $\SM$ as
\begin{align}
\left(\frac{\partial \SM}{\partial H}\right)_T = \mu_0\left(\frac{\partial M}{\partial T}\right)_H.
\end{align}
Therefore, we can estimate $(\partial \SM/\partial H)_T$ from $M$ reported by Tenya \etal~\cite{Tenya2006.JPhysSocJpn.75.023702.SM}. 
Because they reported $M(H)$ curves only at a few fixed temperatures, we estimated $(\partial \SM/\partial H)_T$ as follows:
We first extracted $M(T)$ data from Fig.~2 of Ref.~\cite{Tenya2006.JPhysSocJpn.75.023702.SM}, and then fitted a quadratic function $M(T)=M_0 + AT^2$ to the data. 
This fitting was successful only below 1.1~T; and note that, even below 1.1~T, only 3-5 data points are available for the fitting.
Because $(\partial M/\partial T)_H \sim 2AT$, $(\partial \SM/\partial H)_T$ divided by $\mu_0T$ is given by $2A$, which is plotted in Fig.~\ref{fig:entropy-compare}(b) after an appropriate conversion of the unit. 
In spite of the uncertainty due to the small number of data for the fitting, $(\partial \SM/\partial H)_T$ and $(\partial\SMCE/\partial H)_T$ reasonably agree with each other.

\section{Summary}

To summarize this Supplementary Material, we explain our careful entropy evaluation process in detail and demonstrate that the process is valid although the $\Ts$ variation due to the MCE is less than 1\%.
We emphasize that a large MCE signal ($\varDelta\Ts(H)$) may not necessarily mean that the evaluated entropy is accurate. 
As temperature variation due to the MCE becomes larger, a violation of the constant-temperature condition becomes much serious.
Then, for a substantially large MCE signal, 
evaluation of $S(H)$ would become complicated because it leads to a large error if one evaluates $S(H)$ by integrating $\dSdH$ from a single measurement curve.
In contrast, in our approach,  we can safely integrate the experimentally-obtained $\dSdH$ to obtain $S(H)$ because the constant-temperature conditions is almost satisfied.
Therefore, our approach provides an alternative route to evaluate the entropy from MCE results.

\appendix*

\renewcommand{\theequation}{SA\arabic{equation}}
\setcounter{equation}{0}
\renewcommand{\thefigure}{SA\arabic{figure}}
\setcounter{figure}{0}
\renewcommand{\thetable}{SA\arabic{table}}
\setcounter{table}{0}

\section{Details of the estimation of reading errors in thermometry}\label{sec:app:error}

Readings of the measured sample temperature $\Tsr$ inevitably contains reading errors $\delta \Ts$ with respect to the true temperature $\Ts\spsm{true}$:
\begin{align}
\Tsr = \Ts\spsm{true} + \delta \Ts.
\label{eq:error}
\end{align}
This is also the case for $\Tsz$, $\Tb$ and $\Tbz$.
We here evaluate influence of these errors to the modified definition $\Dt = [(\Ts/\Tsz)-(\Tb/\Tbz)]/(\Ts/\Tsz)$. Putting \eq{eq:error} to the definition of $\Dt$ (\eq{eq:t-definition}), the experimental value of $\Dt$ can be written as
\begin{align}
\Dt & = 1 -\frac{\Tb}{\Tbz}\frac{\Tsz}{\Ts},\\
    & = 1-\frac{\Tb\spsm{true}+\delta\Tb}{\Tbz\spsm{true}+\delta\Tbz}\frac{\Tsz\spsm{true}+\delta\Tsz}{\Ts\spsm{true}+\delta\Ts}, \\
& \simeq 1 -  \frac{\Tb\spsm{true}}{\Tbz\spsm{true}}\frac{\Tsz\spsm{true}}{\Ts\spsm{true}}\left(1 + \frac{\delta\Tb}{\Tb\spsm{true}} - \frac{\delta\Tbz}{\Tbz\spsm{true}} - \frac{\delta\Ts}{\Ts\spsm{true}} + \frac{\delta\Tsz}{\Tsz\spsm{true}}\right).\label{eq:error2}
\end{align}
Here, we neglect second order terms of the small values. 
Because of the approximation
\begin{align*}
\frac{\delta\Ts}{\Ts\spsm{true}}=\frac{\delta\Ts}{\Tsz\spsm{true} + (\Ts\spsm{true}-\Tsz\spsm{true})} \sim \frac{\delta\Ts}{\Tsz\spsm{true}}\left(1-\frac{\Ts\spsm{true}-\Tsz\spsm{true}}{\Tsz\spsm{true}}\right) \sim \frac{\delta\Ts}{\Tsz\spsm{true}}
\end{align*}
within the first order to the small values (such as $\delta\Ts/\Tsz\spsm{true}$ and $(\Ts\spsm{true}-\Tsz\spsm{true})/\Tsz\spsm{true}$), and because of similar approximations for $\Tb$ and $\Tbz$, 
we can substitute the denominators in the parentheses in \eq{eq:error2} by $\Tsz\spsm{true}$ as
\begin{align}
\Dt \simeq 1 -  \frac{\Tb\spsm{true}}{\Tbz\spsm{true}}\frac{\Tsz\spsm{true}}{\Ts\spsm{true}}\left(1 + \frac{\delta\Tb - \delta\Tbz - \delta\Ts +  \delta\Tsz}{\Tsz\spsm{true}}\right).\label{eq:error3}
\end{align}

The reading errors $\delta T$ can be separated into a term originating from calibration errors and a term originating from electronics drift:
\begin{align}
\delta T = \delta T\spsm{cal}(T,H) + \delta T\spsm{drift}.
\end{align}
Note that we can use the same function $\delta T\spsm{cal}(T,H)$ for all four temperatures discussed here because they are measured with the same thermometer.
In addition, the drift term should satisfy $\delta\Ts\spsm{drift}=\delta\Tsz\spsm{drift}$ and $\delta\Tb\spsm{drift}=\delta\Tbz\spsm{drift}$.
Recalling the fact that $\Tsz$ and $\Tbz$ are the values at a certain field $H = H_0$, we can write the reading errors as
\begin{align}
\delta \Ts &= \delta T\spsm{cal}(\Ts\spsm{true},H) + \delta \Ts\spsm{drift},\\
\delta \Tsz &= \delta T\spsm{cal}(\Tsz\spsm{true},H_0) + \delta \Ts\spsm{drift},\\
\delta \Tb &= \delta T\spsm{cal}(\Tb\spsm{true},H) + \delta \Tb\spsm{drift},\\
\delta \Tbz &= \delta T\spsm{cal}(\Tbz\spsm{true},H_0) + \delta \Tb\spsm{drift}.
\end{align}
The temperature variation in the calibration error $\delta T\spsm{cal}$ should be small in the present small temperature range, i.e. $\delta T\spsm{cal}(\Ts\spsm{true},H) \simeq \delta T\spsm{cal}(\Tb\spsm{true},H)$ and $\delta T\spsm{cal}(\Tsz\spsm{true},H_0) \simeq \delta T\spsm{cal}(\Tbz\spsm{true},H_0)$. Corrections to this simplification only results in higher order errors. Thus, we can conclude 
\begin{align}
\delta\Tb - \delta\Tbz - \delta\Ts +  \delta\Tsz \simeq 0
\end{align}
and
\begin{align}
\Dt \simeq 1 -  \frac{\Tb\spsm{true}}{\Tbz\spsm{true}}\frac{\Tsz\spsm{true}}{\Ts\spsm{true}} = 1-\frac{\Tb\spsm{true}}{\Ts\spsm{true}}\,\,\,(\because \Tsz\spsm{true} = \Tbz\spsm{true})
\end{align}
up to the first order in small values.

To summarize, use of the modified definition $\Dt = [(\Ts/\Tsz)-(\Tb/\Tbz)]/(\Ts/\Tsz)$ minimizes influences of the reading errors to the entropy evaluation.


\begin{thebibliography}{44}%
\makeatletter
\providecommand \@ifxundefined [1]{%
 \@ifx{#1\undefined}
}%
\providecommand \@ifnum [1]{%
 \ifnum #1\expandafter \@firstoftwo
 \else \expandafter \@secondoftwo
 \fi
}%
\providecommand \@ifx [1]{%
 \ifx #1\expandafter \@firstoftwo
 \else \expandafter \@secondoftwo
 \fi
}%
\providecommand \natexlab [1]{#1}%
\providecommand \enquote  [1]{``#1''}%
\providecommand \bibnamefont  [1]{#1}%
\providecommand \bibfnamefont [1]{#1}%
\providecommand \citenamefont [1]{#1}%
\providecommand \href@noop [0]{\@secondoftwo}%
\providecommand \href [0]{\begingroup \@sanitize@url \@href}%
\providecommand \@href[1]{\@@startlink{#1}\@@href}%
\providecommand \@@href[1]{\endgroup#1\@@endlink}%
\providecommand \@sanitize@url [0]{\catcode `\\12\catcode `\$12\catcode
  `\&12\catcode `\#12\catcode `\^12\catcode `\_12\catcode `\%12\relax}%
\providecommand \@@startlink[1]{}%
\providecommand \@@endlink[0]{}%
\providecommand \url  [0]{\begingroup\@sanitize@url \@url }%
\providecommand \@url [1]{\endgroup\@href {#1}{\urlprefix }}%
\providecommand \urlprefix  [0]{URL }%
\providecommand \Eprint [0]{\href }%
\@ifxundefined \urlstyle {%
  \providecommand \doi  [0]{\begingroup \@sanitize@url \@doi}%
  \providecommand \@doi [1]{\endgroup \@@startlink {\doibase
  #1}doi:\discretionary {}{}{}#1\@@endlink }%
}{%
  \providecommand \doi  [0]{doi:\discretionary{}{}{}\begingroup
  \urlstyle{rm}\Url }%
}%
\providecommand \doibase [0]{http://dx.doi.org/}%
\providecommand \Doi [0]{\begingroup \@sanitize@url \@Doi }%
\providecommand \@Doi  [1]{\endgroup\@@startlink{\doibase#1}\@@Doi}%
\providecommand \@@Doi [1]{#1\@@endlink}%
\providecommand \selectlanguage [0]{\@gobble}%
\providecommand \bibinfo  [0]{\@secondoftwo}%
\providecommand \bibfield  [0]{\@secondoftwo}%
\providecommand \translation [1]{[#1]}%
\providecommand \BibitemOpen [0]{}%
\providecommand \bibitemStop [0]{}%
\providecommand \bibitemNoStop [0]{.\EOS\space}%
\providecommand \EOS [0]{\spacefactor3000\relax}%
\providecommand \BibitemShut  [1]{\csname bibitem#1\endcsname}%
\bibitem [{\citenamefont {Tinkham}(1996)}]{TinkhamText}%
  \BibitemOpen
  \bibfield  {author} {\bibinfo {author} {\bibfnamefont {M.}~\bibnamefont
  {Tinkham}},\ }\href@noop {} {\emph {\bibinfo {title} {Introduction to
  Superconductivity, Second Edition}}}\ (\bibinfo  {publisher} {McGraw-Hill},\
  \bibinfo {address} {New York},\ \bibinfo {year} {1996})\BibitemShut {NoStop}%
\bibitem [{\citenamefont {Clogston}(1962)}]{Clogston1962}%
  \BibitemOpen
  \bibfield  {author} {\bibinfo {author} {\bibfnamefont {A.~M.}\ \bibnamefont
  {Clogston}},\ }\href@noop {} {\bibfield  {journal} {\bibinfo  {journal}
  {Phys. Rev. Lett.}\ }\textbf {\bibinfo {volume} {9}},\ \bibinfo {pages} {266}
  (\bibinfo {year} {1962})}\BibitemShut {NoStop}%
\bibitem [{\citenamefont {Matsuda}\ and\ \citenamefont
  {Shimahara}(2007)}]{Matsuda2007JPhysSocJpnReview}%
  \BibitemOpen
  \bibfield  {author} {\bibinfo {author} {\bibfnamefont {Y.}~\bibnamefont
  {Matsuda}}\ and\ \bibinfo {author} {\bibfnamefont {H.}~\bibnamefont
  {Shimahara}},\ }\Doi {10.1143/JPSJ.76.051005} {\bibfield  {journal} {\bibinfo
   {journal} {J. Phys. Soc. Jpn.}\ }\textbf {\bibinfo {volume} {76}},\ \bibinfo
  {pages} {051005} (\bibinfo {year} {2007})},\ \bibinfo {note} {and references
  therein.}\BibitemShut {Stop}%
\bibitem [{\citenamefont {Bianchi}\ \emph {et~al.}(2002)\citenamefont
  {Bianchi}, \citenamefont {Movshovich}, \citenamefont {Oeschler},
  \citenamefont {Gegenwart}, \citenamefont {Steglich}, \citenamefont
  {Thompson}, \citenamefont {Pagliuso},\ and\ \citenamefont
  {Sarrao}}]{Bianchi2002}%
  \BibitemOpen
  \bibfield  {author} {\bibinfo {author} {\bibfnamefont {A.}~\bibnamefont
  {Bianchi}}, \bibinfo {author} {\bibfnamefont {R.}~\bibnamefont {Movshovich}},
  \bibinfo {author} {\bibfnamefont {N.}~\bibnamefont {Oeschler}}, \bibinfo
  {author} {\bibfnamefont {P.}~\bibnamefont {Gegenwart}}, \bibinfo {author}
  {\bibfnamefont {F.}~\bibnamefont {Steglich}}, \bibinfo {author}
  {\bibfnamefont {J.~D.}\ \bibnamefont {Thompson}}, \bibinfo {author}
  {\bibfnamefont {P.~G.}\ \bibnamefont {Pagliuso}},\ and\ \bibinfo {author}
  {\bibfnamefont {J.~L.}\ \bibnamefont {Sarrao}},\ }\href@noop {} {\bibfield
  {journal} {\bibinfo  {journal} {Phys. Rev. Lett.}\ }\textbf {\bibinfo
  {volume} {89}},\ \bibinfo {pages} {137002} (\bibinfo {year}
  {2002})}\BibitemShut {NoStop}%
\bibitem [{\citenamefont {Radovan}\ \emph {et~al.}(2003)\citenamefont
  {Radovan}, \citenamefont {Fortune}, \citenamefont {Murphy}, \citenamefont
  {Hannahs}, \citenamefont {Palm}, \citenamefont {Tozer},\ and\ \citenamefont
  {Hall}}]{Radovan2003.Nature.425.51}%
  \BibitemOpen
  \bibfield  {author} {\bibinfo {author} {\bibfnamefont {H.~A.}\ \bibnamefont
  {Radovan}}, \bibinfo {author} {\bibfnamefont {N.~A.}\ \bibnamefont
  {Fortune}}, \bibinfo {author} {\bibfnamefont {T.~P.}\ \bibnamefont {Murphy}},
  \bibinfo {author} {\bibfnamefont {S.~T.}\ \bibnamefont {Hannahs}}, \bibinfo
  {author} {\bibfnamefont {E.~C.}\ \bibnamefont {Palm}}, \bibinfo {author}
  {\bibfnamefont {S.~W.}\ \bibnamefont {Tozer}},\ and\ \bibinfo {author}
  {\bibfnamefont {D.}~\bibnamefont {Hall}},\ }\Doi {10.1038/nature01842}
  {\bibfield  {journal} {\bibinfo  {journal} {Nature}\ }\textbf {\bibinfo
  {volume} {425}},\ \bibinfo {pages} {51} (\bibinfo {year} {2003})}\BibitemShut
  {NoStop}%
\bibitem [{\citenamefont {Lortz}\ \emph {et~al.}(2007)\citenamefont {Lortz},
  \citenamefont {Wang}, \citenamefont {Demuer}, \citenamefont {Bottger},
  \citenamefont {Bergk}, \citenamefont {Zwicknagl}, \citenamefont {Nakazawa},\
  and\ \citenamefont {Wosnitza}}]{Lortz2007PhysRevLett}%
  \BibitemOpen
  \bibfield  {author} {\bibinfo {author} {\bibfnamefont {R.}~\bibnamefont
  {Lortz}}, \bibinfo {author} {\bibfnamefont {Y.}~\bibnamefont {Wang}},
  \bibinfo {author} {\bibfnamefont {A.}~\bibnamefont {Demuer}}, \bibinfo
  {author} {\bibfnamefont {P.~H.~M.}\ \bibnamefont {Bottger}}, \bibinfo
  {author} {\bibfnamefont {B.}~\bibnamefont {Bergk}}, \bibinfo {author}
  {\bibfnamefont {G.}~\bibnamefont {Zwicknagl}}, \bibinfo {author}
  {\bibfnamefont {Y.}~\bibnamefont {Nakazawa}},\ and\ \bibinfo {author}
  {\bibfnamefont {J.}~\bibnamefont {Wosnitza}},\ }\Doi
  {10.1103/PhysRevLett.99.187002} {\bibfield  {journal} {\bibinfo  {journal}
  {Phys. Rev. Lett.}\ }\textbf {\bibinfo {volume} {99}},\ \bibinfo {pages}
  {187002} (\bibinfo {year} {2007})}\BibitemShut {NoStop}%
\bibitem [{\citenamefont {Maeno}\ \emph {et~al.}(1994)\citenamefont {Maeno},
  \citenamefont {Hashimoto}, \citenamefont {Yoshida}, \citenamefont
  {Nishizaki}, \citenamefont {Fujita}, \citenamefont {Bednorz},\ and\
  \citenamefont {Lichtenberg}}]{Maeno1994}%
  \BibitemOpen
  \bibfield  {author} {\bibinfo {author} {\bibfnamefont {Y.}~\bibnamefont
  {Maeno}}, \bibinfo {author} {\bibfnamefont {H.}~\bibnamefont {Hashimoto}},
  \bibinfo {author} {\bibfnamefont {K.}~\bibnamefont {Yoshida}}, \bibinfo
  {author} {\bibfnamefont {S.}~\bibnamefont {Nishizaki}}, \bibinfo {author}
  {\bibfnamefont {T.}~\bibnamefont {Fujita}}, \bibinfo {author} {\bibfnamefont
  {J.~G.}\ \bibnamefont {Bednorz}},\ and\ \bibinfo {author} {\bibfnamefont
  {F.}~\bibnamefont {Lichtenberg}},\ }\href@noop {} {\bibfield  {journal}
  {\bibinfo  {journal} {Nature}\ }\textbf {\bibinfo {volume} {372}},\ \bibinfo
  {pages} {532} (\bibinfo {year} {1994})}\BibitemShut {NoStop}%
\bibitem [{\citenamefont {Mackenzie}\ and\ \citenamefont
  {Maeno}(2003)}]{Mackenzie2003RMP}%
  \BibitemOpen
  \bibfield  {author} {\bibinfo {author} {\bibfnamefont {A.~P.}\ \bibnamefont
  {Mackenzie}}\ and\ \bibinfo {author} {\bibfnamefont {Y.}~\bibnamefont
  {Maeno}},\ }\href@noop {} {\bibfield  {journal} {\bibinfo  {journal} {Rev.
  Mod. Phys.}\ }\textbf {\bibinfo {volume} {75}},\ \bibinfo {pages} {657}
  (\bibinfo {year} {2003})}\BibitemShut {NoStop}%
\bibitem [{\citenamefont {Maeno}\ \emph {et~al.}(2011)\citenamefont {Maeno},
  \citenamefont {Kittaka}, \citenamefont {Nomura}, \citenamefont {Yonezawa},\
  and\ \citenamefont {Ishida}}]{Maeno2011.JPhysSocJpn.81.011009}%
  \BibitemOpen
  \bibfield  {author} {\bibinfo {author} {\bibfnamefont {Y.}~\bibnamefont
  {Maeno}}, \bibinfo {author} {\bibfnamefont {S.}~\bibnamefont {Kittaka}},
  \bibinfo {author} {\bibfnamefont {T.}~\bibnamefont {Nomura}}, \bibinfo
  {author} {\bibfnamefont {S.}~\bibnamefont {Yonezawa}},\ and\ \bibinfo
  {author} {\bibfnamefont {K.}~\bibnamefont {Ishida}},\ }\Doi
  {10.1143/JPSJ.81.011009} {\bibfield  {journal} {\bibinfo  {journal} {J. Phys.
  Soc. Jpn.}\ }\textbf {\bibinfo {volume} {81}},\ \bibinfo {pages} {011009}
  (\bibinfo {year} {2011})}\BibitemShut {NoStop}%
\bibitem [{\citenamefont {Nelson}\ \emph {et~al.}(2004)\citenamefont {Nelson},
  \citenamefont {Mao}, \citenamefont {Maeno},\ and\ \citenamefont
  {Liu}}]{Nelson2004.Science.306.1151}%
  \BibitemOpen
  \bibfield  {author} {\bibinfo {author} {\bibfnamefont {K.~D.}\ \bibnamefont
  {Nelson}}, \bibinfo {author} {\bibfnamefont {Z.~Q.}\ \bibnamefont {Mao}},
  \bibinfo {author} {\bibfnamefont {Y.}~\bibnamefont {Maeno}},\ and\ \bibinfo
  {author} {\bibfnamefont {Y.}~\bibnamefont {Liu}},\ }\Doi
  {10.1126/science.1103881} {\bibfield  {journal} {\bibinfo  {journal}
  {Science}\ }\textbf {\bibinfo {volume} {306}},\ \bibinfo {pages} {1151}
  (\bibinfo {year} {2004})}\BibitemShut {NoStop}%
\bibitem [{\citenamefont {Xia}\ \emph {et~al.}(2006)\citenamefont {Xia},
  \citenamefont {Maeno}, \citenamefont {Beyersdorf}, \citenamefont {Fejer},\
  and\ \citenamefont {Kapitulnik}}]{Xia2006.PhysRevLett.97.167002}%
  \BibitemOpen
  \bibfield  {author} {\bibinfo {author} {\bibfnamefont {J.}~\bibnamefont
  {Xia}}, \bibinfo {author} {\bibfnamefont {Y.}~\bibnamefont {Maeno}}, \bibinfo
  {author} {\bibfnamefont {P.~T.}\ \bibnamefont {Beyersdorf}}, \bibinfo
  {author} {\bibfnamefont {M.~M.}\ \bibnamefont {Fejer}},\ and\ \bibinfo
  {author} {\bibfnamefont {A.}~\bibnamefont {Kapitulnik}},\ }\Doi
  {10.1103/PhysRevLett.97.167002} {\bibfield  {journal} {\bibinfo  {journal}
  {Phys. Rev. Lett.}\ }\textbf {\bibinfo {volume} {97}},\ \bibinfo {pages}
  {167002} (\bibinfo {year} {2006})}\BibitemShut {NoStop}%
\bibitem [{\citenamefont {Kashiwaya}\ \emph {et~al.}(2011)\citenamefont
  {Kashiwaya}, \citenamefont {Kashiwaya}, \citenamefont {Kambara},
  \citenamefont {Furuta}, \citenamefont {Yaguchi}, \citenamefont {Tanaka},\
  and\ \citenamefont {Maeno}}]{Kashiwaya2011.PhysRevLett.107.077003}%
  \BibitemOpen
  \bibfield  {author} {\bibinfo {author} {\bibfnamefont {S.}~\bibnamefont
  {Kashiwaya}}, \bibinfo {author} {\bibfnamefont {H.}~\bibnamefont
  {Kashiwaya}}, \bibinfo {author} {\bibfnamefont {H.}~\bibnamefont {Kambara}},
  \bibinfo {author} {\bibfnamefont {T.}~\bibnamefont {Furuta}}, \bibinfo
  {author} {\bibfnamefont {H.}~\bibnamefont {Yaguchi}}, \bibinfo {author}
  {\bibfnamefont {Y.}~\bibnamefont {Tanaka}},\ and\ \bibinfo {author}
  {\bibfnamefont {Y.}~\bibnamefont {Maeno}},\ }\Doi
  {10.1103/PhysRevLett.107.077003} {\bibfield  {journal} {\bibinfo  {journal}
  {Phys. Rev. Lett.}\ }\textbf {\bibinfo {volume} {107}},\ \bibinfo {pages}
  {077003} (\bibinfo {year} {2011})}\BibitemShut {NoStop}%
\bibitem [{\citenamefont {Nakamura}\ \emph {et~al.}(2011)\citenamefont
  {Nakamura}, \citenamefont {Nakagawa}, \citenamefont {Yamagishi},
  \citenamefont {Terashima}, \citenamefont {Yonezawa}, \citenamefont
  {Sigrist},\ and\ \citenamefont {Maeno}}]{Nakamura2011.PhysRevB.84.060512R}%
  \BibitemOpen
  \bibfield  {author} {\bibinfo {author} {\bibfnamefont {T.}~\bibnamefont
  {Nakamura}}, \bibinfo {author} {\bibfnamefont {R.}~\bibnamefont {Nakagawa}},
  \bibinfo {author} {\bibfnamefont {T.}~\bibnamefont {Yamagishi}}, \bibinfo
  {author} {\bibfnamefont {T.}~\bibnamefont {Terashima}}, \bibinfo {author}
  {\bibfnamefont {S.}~\bibnamefont {Yonezawa}}, \bibinfo {author}
  {\bibfnamefont {M.}~\bibnamefont {Sigrist}},\ and\ \bibinfo {author}
  {\bibfnamefont {Y.}~\bibnamefont {Maeno}},\ }\Doi
  {10.1103/PhysRevB.84.060512} {\bibfield  {journal} {\bibinfo  {journal}
  {Phys. Rev. B}\ }\textbf {\bibinfo {volume} {84}},\ \bibinfo {pages}
  {060512(R)} (\bibinfo {year} {2011})}\BibitemShut {NoStop}%
\bibitem [{\citenamefont {Jang}\ \emph {et~al.}(2011)\citenamefont {Jang},
  \citenamefont {Ferguson}, \citenamefont {Vakaryuk}, \citenamefont {Budakian},
  \citenamefont {Chung}, \citenamefont {Goldbart},\ and\ \citenamefont
  {Maeno}}]{Jang2011.Science.331.186}%
  \BibitemOpen
  \bibfield  {author} {\bibinfo {author} {\bibfnamefont {J.}~\bibnamefont
  {Jang}}, \bibinfo {author} {\bibfnamefont {D.~G.}\ \bibnamefont {Ferguson}},
  \bibinfo {author} {\bibfnamefont {V.}~\bibnamefont {Vakaryuk}}, \bibinfo
  {author} {\bibfnamefont {R.}~\bibnamefont {Budakian}}, \bibinfo {author}
  {\bibfnamefont {S.~B.}\ \bibnamefont {Chung}}, \bibinfo {author}
  {\bibfnamefont {P.~M.}\ \bibnamefont {Goldbart}},\ and\ \bibinfo {author}
  {\bibfnamefont {Y.}~\bibnamefont {Maeno}},\ }\Doi {10.1126/science.1193839}
  {\bibfield  {journal} {\bibinfo  {journal} {Science}\ }\textbf {\bibinfo
  {volume} {331}},\ \bibinfo {pages} {186} (\bibinfo {year}
  {2011})}\BibitemShut {NoStop}%
\bibitem [{\citenamefont {Ishida}\ \emph {et~al.}(1998)\citenamefont {Ishida},
  \citenamefont {Mukuda}, \citenamefont {Kitaoka}, \citenamefont {Asayama},
  \citenamefont {Mao}, \citenamefont {Mori},\ and\ \citenamefont
  {Maeno}}]{Ishida1998.Nature.396.658}%
  \BibitemOpen
  \bibfield  {author} {\bibinfo {author} {\bibfnamefont {K.}~\bibnamefont
  {Ishida}}, \bibinfo {author} {\bibfnamefont {H.}~\bibnamefont {Mukuda}},
  \bibinfo {author} {\bibfnamefont {Y.}~\bibnamefont {Kitaoka}}, \bibinfo
  {author} {\bibfnamefont {K.}~\bibnamefont {Asayama}}, \bibinfo {author}
  {\bibfnamefont {Z.~Q.}\ \bibnamefont {Mao}}, \bibinfo {author} {\bibfnamefont
  {Y.}~\bibnamefont {Mori}},\ and\ \bibinfo {author} {\bibfnamefont
  {Y.}~\bibnamefont {Maeno}},\ }\Doi {10.1038/25315} {\bibfield  {journal}
  {\bibinfo  {journal} {Nature}\ }\textbf {\bibinfo {volume} {396}},\ \bibinfo
  {pages} {658} (\bibinfo {year} {1998})}\BibitemShut {NoStop}%
\bibitem [{\citenamefont {Ishida}\ \emph {et~al.}(2001)\citenamefont {Ishida},
  \citenamefont {Mukuda}, \citenamefont {Kitaoka}, \citenamefont {Mao},
  \citenamefont {Fukazawa},\ and\ \citenamefont
  {Maeno}}]{Ishida2001.PhysRevB.63.060507R}%
  \BibitemOpen
  \bibfield  {author} {\bibinfo {author} {\bibfnamefont {K.}~\bibnamefont
  {Ishida}}, \bibinfo {author} {\bibfnamefont {H.}~\bibnamefont {Mukuda}},
  \bibinfo {author} {\bibfnamefont {Y.}~\bibnamefont {Kitaoka}}, \bibinfo
  {author} {\bibfnamefont {Z.~Q.}\ \bibnamefont {Mao}}, \bibinfo {author}
  {\bibfnamefont {H.}~\bibnamefont {Fukazawa}},\ and\ \bibinfo {author}
  {\bibfnamefont {Y.}~\bibnamefont {Maeno}},\ }\Doi
  {10.1103/PhysRevB.63.060507} {\bibfield  {journal} {\bibinfo  {journal}
  {Phys. Rev. B}\ }\textbf {\bibinfo {volume} {63}},\ \bibinfo {pages}
  {060507(R)} (\bibinfo {year} {2001})}\BibitemShut {NoStop}%
\bibitem [{\citenamefont {Murakawa}\ \emph {et~al.}(2004)\citenamefont
  {Murakawa}, \citenamefont {Ishida}, \citenamefont {Kitagawa}, \citenamefont
  {Mao},\ and\ \citenamefont {Maeno}}]{Murakawa2004.PhysRevLett.93.167004}%
  \BibitemOpen
  \bibfield  {author} {\bibinfo {author} {\bibfnamefont {H.}~\bibnamefont
  {Murakawa}}, \bibinfo {author} {\bibfnamefont {K.}~\bibnamefont {Ishida}},
  \bibinfo {author} {\bibfnamefont {K.}~\bibnamefont {Kitagawa}}, \bibinfo
  {author} {\bibfnamefont {Z.~Q.}\ \bibnamefont {Mao}},\ and\ \bibinfo {author}
  {\bibfnamefont {Y.}~\bibnamefont {Maeno}},\ }\Doi
  {10.1103/PhysRevLett.93.167004} {\bibfield  {journal} {\bibinfo  {journal}
  {Phys. Rev. Lett.}\ }\textbf {\bibinfo {volume} {93}},\ \bibinfo {pages}
  {167004} (\bibinfo {year} {2004})}\BibitemShut {NoStop}%
\bibitem [{\citenamefont {Duffy}\ \emph {et~al.}(2000)\citenamefont {Duffy},
  \citenamefont {Hayden}, \citenamefont {Maeno}, \citenamefont {Mao},
  \citenamefont {Kulda},\ and\ \citenamefont
  {McIntyre}}]{Duffy2000.PhysRevLett.85.5412}%
  \BibitemOpen
  \bibfield  {author} {\bibinfo {author} {\bibfnamefont {J.~A.}\ \bibnamefont
  {Duffy}}, \bibinfo {author} {\bibfnamefont {S.~M.}\ \bibnamefont {Hayden}},
  \bibinfo {author} {\bibfnamefont {Y.}~\bibnamefont {Maeno}}, \bibinfo
  {author} {\bibfnamefont {Z.}~\bibnamefont {Mao}}, \bibinfo {author}
  {\bibfnamefont {J.}~\bibnamefont {Kulda}},\ and\ \bibinfo {author}
  {\bibfnamefont {G.~J.}\ \bibnamefont {McIntyre}},\ }\Doi
  {10.1103/PhysRevLett.85.5412} {\bibfield  {journal} {\bibinfo  {journal}
  {Phys. Rev. Lett.}\ }\textbf {\bibinfo {volume} {85}},\ \bibinfo {pages}
  {5412} (\bibinfo {year} {2000})}\BibitemShut {NoStop}%
\bibitem [{\citenamefont {Akima}\ \emph {et~al.}(1999)\citenamefont {Akima},
  \citenamefont {Nishizaki},\ and\ \citenamefont
  {Maeno}}]{Akima1999.JPhysSocJpn.68.694}%
  \BibitemOpen
  \bibfield  {author} {\bibinfo {author} {\bibfnamefont {T.}~\bibnamefont
  {Akima}}, \bibinfo {author} {\bibfnamefont {S.}~\bibnamefont {Nishizaki}},\
  and\ \bibinfo {author} {\bibfnamefont {Y.}~\bibnamefont {Maeno}},\ }\Doi
  {10.1143/JPSJ.68.694} {\bibfield  {journal} {\bibinfo  {journal} {J. Phys.
  Soc. Jpn.}\ }\textbf {\bibinfo {volume} {68}},\ \bibinfo {pages} {694}
  (\bibinfo {year} {1999})}\BibitemShut {NoStop}%
\bibitem [{\citenamefont {Deguchi}\ \emph {et~al.}(2002)\citenamefont
  {Deguchi}, \citenamefont {Tanatar}, \citenamefont {Mao}, \citenamefont
  {Ishiguro},\ and\ \citenamefont {Maeno}}]{Deguchi2002}%
  \BibitemOpen
  \bibfield  {author} {\bibinfo {author} {\bibfnamefont {K.}~\bibnamefont
  {Deguchi}}, \bibinfo {author} {\bibfnamefont {M.~A.}\ \bibnamefont
  {Tanatar}}, \bibinfo {author} {\bibfnamefont {Z.}~\bibnamefont {Mao}},
  \bibinfo {author} {\bibfnamefont {T.}~\bibnamefont {Ishiguro}},\ and\
  \bibinfo {author} {\bibfnamefont {Y.}~\bibnamefont {Maeno}},\ }\href@noop {}
  {\bibfield  {journal} {\bibinfo  {journal} {J. Phys. Soc. Jpn.}\ }\textbf
  {\bibinfo {volume} {71}},\ \bibinfo {pages} {2839} (\bibinfo {year}
  {2002})}\BibitemShut {NoStop}%
\bibitem [{\citenamefont {Kittaka}\ \emph {et~al.}(2009)\citenamefont
  {Kittaka}, \citenamefont {Nakamura}, \citenamefont {Aono}, \citenamefont
  {Yonezawa}, \citenamefont {Ishida},\ and\ \citenamefont
  {Maeno}}]{Kittaka2009.PhysRevB.80.174514}%
  \BibitemOpen
  \bibfield  {author} {\bibinfo {author} {\bibfnamefont {S.}~\bibnamefont
  {Kittaka}}, \bibinfo {author} {\bibfnamefont {T.}~\bibnamefont {Nakamura}},
  \bibinfo {author} {\bibfnamefont {Y.}~\bibnamefont {Aono}}, \bibinfo {author}
  {\bibfnamefont {S.}~\bibnamefont {Yonezawa}}, \bibinfo {author}
  {\bibfnamefont {K.}~\bibnamefont {Ishida}},\ and\ \bibinfo {author}
  {\bibfnamefont {Y.}~\bibnamefont {Maeno}},\ }\Doi
  {10.1103/PhysRevB.80.174514} {\bibfield  {journal} {\bibinfo  {journal}
  {Phys. Rev. B}\ }\textbf {\bibinfo {volume} {80}},\ \bibinfo {pages} {174514}
  (\bibinfo {year} {2009})}\BibitemShut {NoStop}%
\bibitem [{\citenamefont {Tenya}\ \emph {et~al.}(2006)\citenamefont {Tenya},
  \citenamefont {Yasuda}, \citenamefont {Yokoyama}, \citenamefont {Amitsuka},
  \citenamefont {Deguchi},\ and\ \citenamefont
  {Maeno}}]{Tenya2006.JPhysSocJpn.75.023702}%
  \BibitemOpen
  \bibfield  {author} {\bibinfo {author} {\bibfnamefont {K.}~\bibnamefont
  {Tenya}}, \bibinfo {author} {\bibfnamefont {S.}~\bibnamefont {Yasuda}},
  \bibinfo {author} {\bibfnamefont {M.}~\bibnamefont {Yokoyama}}, \bibinfo
  {author} {\bibfnamefont {H.}~\bibnamefont {Amitsuka}}, \bibinfo {author}
  {\bibfnamefont {K.}~\bibnamefont {Deguchi}},\ and\ \bibinfo {author}
  {\bibfnamefont {Y.}~\bibnamefont {Maeno}},\ }\Doi {0.1143/JPSJ.75.023702}
  {\bibfield  {journal} {\bibinfo  {journal} {J. Phys. Soc. Jpn.}\ }\textbf
  {\bibinfo {volume} {75}},\ \bibinfo {pages} {023702} (\bibinfo {year}
  {2006})}\BibitemShut {NoStop}%
\bibitem [{\citenamefont {Rost}\ \emph {et~al.}(2009)\citenamefont {Rost},
  \citenamefont {Perry}, \citenamefont {Mercure}, \citenamefont {Mackenzie},\
  and\ \citenamefont {Grigera}}]{Rost2009.Science.325.1360}%
  \BibitemOpen
  \bibfield  {author} {\bibinfo {author} {\bibfnamefont {A.~W.}\ \bibnamefont
  {Rost}}, \bibinfo {author} {\bibfnamefont {R.~S.}\ \bibnamefont {Perry}},
  \bibinfo {author} {\bibfnamefont {J.-F.}\ \bibnamefont {Mercure}}, \bibinfo
  {author} {\bibfnamefont {A.~P.}\ \bibnamefont {Mackenzie}},\ and\ \bibinfo
  {author} {\bibfnamefont {S.~A.}\ \bibnamefont {Grigera}},\ }\Doi
  {10.1126/science.1176627} {\bibfield  {journal} {\bibinfo  {journal}
  {Science}\ }\textbf {\bibinfo {volume} {325}},\ \bibinfo {pages} {1360}
  (\bibinfo {year} {2009})}\BibitemShut {NoStop}%
\bibitem [{Not(){\natexlab{a}}}]{NoteSuppleMat}%
  \BibitemOpen
  \bibinfo {note} {{See Supplemental Material attached in the end of this paper for the
  information on the experimental method, the entropy evaluation, and the
  comparison of the obtained entropy with previous studies.}}\BibitemShut
  {Stop}%
\bibitem [{\citenamefont {Mao}\ \emph {et~al.}(2000)\citenamefont {Mao},
  \citenamefont {Maeno},\ and\ \citenamefont
  {Fukazawa}}]{Mao2000.MaterResBull.35.1813}%
  \BibitemOpen
  \bibfield  {author} {\bibinfo {author} {\bibfnamefont {Z.}~\bibnamefont
  {Mao}}, \bibinfo {author} {\bibfnamefont {Y.}~\bibnamefont {Maeno}},\ and\
  \bibinfo {author} {\bibfnamefont {H.}~\bibnamefont {Fukazawa}},\ }\Doi
  {10.1016/S0025-5408(00)00378-0} {\bibfield  {journal} {\bibinfo  {journal}
  {Mater. Res. Bull.}\ }\textbf {\bibinfo {volume} {35}},\ \bibinfo {pages}
  {1813} (\bibinfo {year} {2000})}\BibitemShut {NoStop}%
\bibitem [{\citenamefont {Mackenzie}\ \emph {et~al.}(1998)\citenamefont
  {Mackenzie}, \citenamefont {Haselwimmer}, \citenamefont {Tyler},
  \citenamefont {Lonzarich}, \citenamefont {Mori}, \citenamefont {Nishizaki},\
  and\ \citenamefont {Maeno}}]{Mackenzie1998.PhysRevLett.80.161}%
  \BibitemOpen
  \bibfield  {author} {\bibinfo {author} {\bibfnamefont {A.~P.}\ \bibnamefont
  {Mackenzie}}, \bibinfo {author} {\bibfnamefont {R.~K.~W.}\ \bibnamefont
  {Haselwimmer}}, \bibinfo {author} {\bibfnamefont {A.~W.}\ \bibnamefont
  {Tyler}}, \bibinfo {author} {\bibfnamefont {G.~G.}\ \bibnamefont
  {Lonzarich}}, \bibinfo {author} {\bibfnamefont {Y.}~\bibnamefont {Mori}},
  \bibinfo {author} {\bibfnamefont {S.}~\bibnamefont {Nishizaki}},\ and\
  \bibinfo {author} {\bibfnamefont {Y.}~\bibnamefont {Maeno}},\ }\Doi
  {10.1103/PhysRevLett.80.161} {\bibfield  {journal} {\bibinfo  {journal}
  {Phys. Rev. Lett.}\ }\textbf {\bibinfo {volume} {80}},\ \bibinfo {pages}
  {161} (\bibinfo {year} {1998})}\BibitemShut {NoStop}%
\bibitem [{\citenamefont {Deguchi}\ \emph {et~al.}(2004)\citenamefont
  {Deguchi}, \citenamefont {Ishiguro},\ and\ \citenamefont
  {Maeno}}]{Deguchi2004RSI}%
  \BibitemOpen
  \bibfield  {author} {\bibinfo {author} {\bibfnamefont {K.}~\bibnamefont
  {Deguchi}}, \bibinfo {author} {\bibfnamefont {T.}~\bibnamefont {Ishiguro}},\
  and\ \bibinfo {author} {\bibfnamefont {Y.}~\bibnamefont {Maeno}},\
  }\href@noop {} {\bibfield  {journal} {\bibinfo  {journal} {Rev. Sci.
  Instrum.}\ }\textbf {\bibinfo {volume} {75}},\ \bibinfo {pages} {1188}
  (\bibinfo {year} {2004})}\BibitemShut {NoStop}%
\bibitem [{Not(){\natexlab{b}}}]{NoteAsymmetry}%
  \BibitemOpen
  \bibinfo {note} {{Systematic evolution of the asymmetric component with
  respect to the temperature and field direction supports our scenario that the
  asymmetric component is dominated by energy dissipation in the samples.
  Nevertheless, at present, we cannot deny small contributions of extrinsic
  origins (such as a tiny error in the temperature measurement).}}\BibitemShut
  {Stop}%
\bibitem [{Not(){\natexlab{c}}}]{NoteThermalRelaxationTime}%
  \BibitemOpen
  \bibinfo {note} {{The thermal relaxation time is approximately 3~sec for
  Sample~\#1, and the delay time of the electronics is chosen to be smaller
  than 1~sec. The overall delay time of our equipments is at most 5~sec for
  Sample~\#1 and even shorter for Sample~\#2.}}\BibitemShut {Stop}%
\bibitem [{\citenamefont {NishiZaki}\ \emph {et~al.}(2000)\citenamefont
  {NishiZaki}, \citenamefont {Maeno},\ and\ \citenamefont
  {Mao}}]{NishiZaki2000JPhysSocJpn}%
  \BibitemOpen
  \bibfield  {author} {\bibinfo {author} {\bibfnamefont {S.}~\bibnamefont
  {NishiZaki}}, \bibinfo {author} {\bibfnamefont {Y.}~\bibnamefont {Maeno}},\
  and\ \bibinfo {author} {\bibfnamefont {Z.}~\bibnamefont {Mao}},\ }\Doi
  {10.1143/JPSJ.69.572} {\bibfield  {journal} {\bibinfo  {journal} {J. Phys.
  Soc. Jpn.}\ }\textbf {\bibinfo {volume} {69}},\ \bibinfo {pages} {572}
  (\bibinfo {year} {2000})}\BibitemShut {NoStop}%
\bibitem [{Not(){\natexlab{d}}}]{NoteMagnetization}%
  \BibitemOpen
  \bibinfo {note} {{Here we linearly interpolated $\delta M$ at 0.14~K
  ($\sim-0.015$~emu/g) and 0.41~K ($\sim-0.010$~emu/g) reported in
  Ref.\cite{Tenya2006.JPhysSocJpn.75.023702}.}}\BibitemShut {Stop}%
\bibitem [{\citenamefont {Ando}\ \emph {et~al.}(1999)\citenamefont {Ando},
  \citenamefont {Akima}, \citenamefont {Mori},\ and\ \citenamefont
  {Maeno}}]{Ando1999.JPhysSocJpn.68.1651}%
  \BibitemOpen
  \bibfield  {author} {\bibinfo {author} {\bibfnamefont {T.}~\bibnamefont
  {Ando}}, \bibinfo {author} {\bibfnamefont {T.}~\bibnamefont {Akima}},
  \bibinfo {author} {\bibfnamefont {Y.}~\bibnamefont {Mori}},\ and\ \bibinfo
  {author} {\bibfnamefont {Y.}~\bibnamefont {Maeno}},\ }\Doi
  {10.1143/JPSJ.68.1651} {\bibfield  {journal} {\bibinfo  {journal} {J. Phys.
  Soc. Jpn.}\ }\textbf {\bibinfo {volume} {68}},\ \bibinfo {pages} {1651}
  (\bibinfo {year} {1999})}\BibitemShut {NoStop}%
\bibitem [{\citenamefont {Yaguchi}\ \emph {et~al.}(2003)\citenamefont
  {Yaguchi}, \citenamefont {Wada}, \citenamefont {Akima}, \citenamefont
  {Maeno},\ and\ \citenamefont {Ishiguro}}]{Yaguchi2003.PhysRevB.67.214519}%
  \BibitemOpen
  \bibfield  {author} {\bibinfo {author} {\bibfnamefont {H.}~\bibnamefont
  {Yaguchi}}, \bibinfo {author} {\bibfnamefont {M.}~\bibnamefont {Wada}},
  \bibinfo {author} {\bibfnamefont {T.}~\bibnamefont {Akima}}, \bibinfo
  {author} {\bibfnamefont {Y.}~\bibnamefont {Maeno}},\ and\ \bibinfo {author}
  {\bibfnamefont {T.}~\bibnamefont {Ishiguro}},\ }\Doi
  {10.1103/PhysRevB.67.214519} {\bibfield  {journal} {\bibinfo  {journal}
  {Phys. Rev. B}\ }\textbf {\bibinfo {volume} {67}},\ \bibinfo {pages} {214519}
  (\bibinfo {year} {2003})}\BibitemShut {NoStop}%
\bibitem [{\citenamefont {Machida}\ and\ \citenamefont
  {Ichioka}(2008)}]{Machida2008.PhysRevB.77.184515}%
  \BibitemOpen
  \bibfield  {author} {\bibinfo {author} {\bibfnamefont {K.}~\bibnamefont
  {Machida}}\ and\ \bibinfo {author} {\bibfnamefont {M.}~\bibnamefont
  {Ichioka}},\ }\Doi {10.1103/PhysRevB.77.184515} {\bibfield  {journal}
  {\bibinfo  {journal} {Phys. Rev. B}\ }\textbf {\bibinfo {volume} {77}},\
  \bibinfo {pages} {184515} (\bibinfo {year} {2008})}\BibitemShut {NoStop}%
\bibitem [{Not(){\natexlab{e}}}]{NoteCondensationEnergy}%
  \BibitemOpen
  \bibinfo {note} {{Here, we used the value $\chis=\chin \sim 0.9\times
  10^{-3}$~emu/mol~\cite{Mackenzie2003RMP}. The value of $\Econd$ is obtained
  from the relation $\Econd = (1/2)\mu_0\Hc^2$ with the thermodynamic critical
  field $\mu_0\Hc=0.0194$~T~\cite{Akima1999.JPhysSocJpn.68.694}.}}\BibitemShut
  {Stop}%
\bibitem [{\citenamefont
  {Agterberg}(1998)}]{Agterberg1998.PhysRevLett.80.5184}%
  \BibitemOpen
  \bibfield  {author} {\bibinfo {author} {\bibfnamefont {D.~F.}\ \bibnamefont
  {Agterberg}},\ }\Doi {10.1103/PhysRevLett.80.5184} {\bibfield  {journal}
  {\bibinfo  {journal} {Phys. Rev. Lett.}\ }\textbf {\bibinfo {volume} {80}},\
  \bibinfo {pages} {5184} (\bibinfo {year} {1998})}\BibitemShut {NoStop}%
\bibitem [{\citenamefont {Udagawa}\ \emph {et~al.}(2005)\citenamefont
  {Udagawa}, \citenamefont {Yanase},\ and\ \citenamefont
  {Ogata}}]{UdagawaM2005.PhysRevB.71.024511}%
  \BibitemOpen
  \bibfield  {author} {\bibinfo {author} {\bibfnamefont {M.}~\bibnamefont
  {Udagawa}}, \bibinfo {author} {\bibfnamefont {Y.}~\bibnamefont {Yanase}},\
  and\ \bibinfo {author} {\bibfnamefont {M.}~\bibnamefont {Ogata}},\ }\Doi
  {10.1103/PhysRevB.71.024511} {\bibfield  {journal} {\bibinfo  {journal}
  {Phys. Rev. B}\ }\textbf {\bibinfo {volume} {71}},\ \bibinfo {pages} {024511}
  (\bibinfo {year} {2005})}\BibitemShut {NoStop}%
\bibitem [{\citenamefont {Kaur}\ \emph {et~al.}(2005)\citenamefont {Kaur},
  \citenamefont {Agterberg},\ and\ \citenamefont
  {Kusunose}}]{Kaur2005.PhysRevB.72.144528}%
  \BibitemOpen
  \bibfield  {author} {\bibinfo {author} {\bibfnamefont {R.~P.}\ \bibnamefont
  {Kaur}}, \bibinfo {author} {\bibfnamefont {D.~F.}\ \bibnamefont
  {Agterberg}},\ and\ \bibinfo {author} {\bibfnamefont {H.}~\bibnamefont
  {Kusunose}},\ }\Doi {10.1103/PhysRevB.72.144528} {\bibfield  {journal}
  {\bibinfo  {journal} {Phys. Rev. B}\ }\textbf {\bibinfo {volume} {72}},\
  \bibinfo {pages} {144528} (\bibinfo {year} {2005})}\BibitemShut {NoStop}%
\bibitem [{\citenamefont {Mineev}(2008)}]{Mineev2008.PhysRevB.77.064519}%
  \BibitemOpen
  \bibfield  {author} {\bibinfo {author} {\bibfnamefont {V.~P.}\ \bibnamefont
  {Mineev}},\ }\Doi {10.1103/PhysRevB.77.064519} {\bibfield  {journal}
  {\bibinfo  {journal} {Phys. Rev. B}\ }\textbf {\bibinfo {volume} {77}},\
  \bibinfo {pages} {064519} (\bibinfo {year} {2008})}\BibitemShut {NoStop}%
\bibitem [{\citenamefont {Haverkort}\ \emph {et~al.}(2008)\citenamefont
  {Haverkort}, \citenamefont {Elfimov}, \citenamefont {Tjeng}, \citenamefont
  {Sawatzky},\ and\ \citenamefont
  {Damascelli}}]{Haverkort2008.PhysRevLett.101.026406}%
  \BibitemOpen
  \bibfield  {author} {\bibinfo {author} {\bibfnamefont {M.~W.}\ \bibnamefont
  {Haverkort}}, \bibinfo {author} {\bibfnamefont {I.~S.}\ \bibnamefont
  {Elfimov}}, \bibinfo {author} {\bibfnamefont {L.~H.}\ \bibnamefont {Tjeng}},
  \bibinfo {author} {\bibfnamefont {G.~A.}\ \bibnamefont {Sawatzky}},\ and\
  \bibinfo {author} {\bibfnamefont {A.}~\bibnamefont {Damascelli}},\ }\Doi
  {10.1103/PhysRevLett.101.026406} {\bibfield  {journal} {\bibinfo  {journal}
  {Phys. Rev. Lett.}\ }\textbf {\bibinfo {volume} {101}},\ \bibinfo {pages}
  {026406} (\bibinfo {year} {2008})}\BibitemShut {NoStop}%
\bibitem [{\citenamefont {Iwasawa}\ \emph {et~al.}(2010)\citenamefont
  {Iwasawa}, \citenamefont {Yoshida}, \citenamefont {Hase}, \citenamefont
  {Koikegami}, \citenamefont {Hayashi}, \citenamefont {Jiang}, \citenamefont
  {Shimada}, \citenamefont {Namatame}, \citenamefont {Taniguchi},\ and\
  \citenamefont {Aiura}}]{Iwasawa2010.PhysRevLett.105.226406}%
  \BibitemOpen
  \bibfield  {author} {\bibinfo {author} {\bibfnamefont {H.}~\bibnamefont
  {Iwasawa}}, \bibinfo {author} {\bibfnamefont {Y.}~\bibnamefont {Yoshida}},
  \bibinfo {author} {\bibfnamefont {I.}~\bibnamefont {Hase}}, \bibinfo {author}
  {\bibfnamefont {S.}~\bibnamefont {Koikegami}}, \bibinfo {author}
  {\bibfnamefont {H.}~\bibnamefont {Hayashi}}, \bibinfo {author} {\bibfnamefont
  {J.}~\bibnamefont {Jiang}}, \bibinfo {author} {\bibfnamefont
  {K.}~\bibnamefont {Shimada}}, \bibinfo {author} {\bibfnamefont
  {H.}~\bibnamefont {Namatame}}, \bibinfo {author} {\bibfnamefont
  {M.}~\bibnamefont {Taniguchi}},\ and\ \bibinfo {author} {\bibfnamefont
  {Y.}~\bibnamefont {Aiura}},\ }\Doi {10.1103/PhysRevLett.105.226406}
  {\bibfield  {journal} {\bibinfo  {journal} {Phys. Rev. Lett.}\ }\textbf
  {\bibinfo {volume} {105}},\ \bibinfo {pages} {226406} (\bibinfo {year}
  {2010})}\BibitemShut {NoStop}%
\bibitem [{\citenamefont {Nomura}\ and\ \citenamefont
  {Yamada}(2000)}]{Nomura2000.JPhysSocJpn.69.1856}%
  \BibitemOpen
  \bibfield  {author} {\bibinfo {author} {\bibfnamefont {T.}~\bibnamefont
  {Nomura}}\ and\ \bibinfo {author} {\bibfnamefont {K.}~\bibnamefont
  {Yamada}},\ }\Doi {10.1143/JPSJ.69.1856} {\bibfield  {journal} {\bibinfo
  {journal} {J. Phys. Soc. Jpn.}\ }\textbf {\bibinfo {volume} {69}},\ \bibinfo
  {pages} {1856} (\bibinfo {year} {2000})}\BibitemShut {NoStop}%
\bibitem [{\citenamefont {Luk'yanchuk}\ and\ \citenamefont
  {Mineev}(1986)}]{Lukyanchuk1986.JETPLett.44.233}%
  \BibitemOpen
  \bibfield  {author} {\bibinfo {author} {\bibfnamefont {I.~A.}\ \bibnamefont
  {Luk'yanchuk}}\ and\ \bibinfo {author} {\bibfnamefont {V.~P.}\ \bibnamefont
  {Mineev}},\ }\href@noop {} {\bibfield  {journal} {\bibinfo  {journal} {JETP
  Lett.}\ }\textbf {\bibinfo {volume} {44}},\ \bibinfo {pages} {233} (\bibinfo
  {year} {1986})}\BibitemShut {NoStop}%
\bibitem [{\citenamefont {Vakaryuk}\ and\ \citenamefont
  {Leggett}(2009)}]{Vakaryuk2009.PhysRevLett.103.057003}%
  \BibitemOpen
  \bibfield  {author} {\bibinfo {author} {\bibfnamefont {V.}~\bibnamefont
  {Vakaryuk}}\ and\ \bibinfo {author} {\bibfnamefont {A.~J.}\ \bibnamefont
  {Leggett}},\ }\Doi {10.1103/PhysRevLett.103.057003} {\bibfield  {journal}
  {\bibinfo  {journal} {Phys. Rev. Lett.}\ }\textbf {\bibinfo {volume} {103}},\
  \bibinfo {pages} {057003} (\bibinfo {year} {2009})}\BibitemShut {NoStop}%
\end{thebibliography}

\begin{thebibliography}{1000}%
\makeatletter
\providecommand \@ifxundefined [1]{%
 \@ifx{#1\undefined}
}%
\providecommand \@ifnum [1]{%
 \ifnum #1\expandafter \@firstoftwo
 \else \expandafter \@secondoftwo
 \fi
}%
\providecommand \@ifx [1]{%
 \ifx #1\expandafter \@firstoftwo
 \else \expandafter \@secondoftwo
 \fi
}%
\providecommand \natexlab [1]{#1}%
\providecommand \enquote  [1]{``#1''}%
\providecommand \bibnamefont  [1]{#1}%
\providecommand \bibfnamefont [1]{#1}%
\providecommand \citenamefont [1]{#1}%
\providecommand \href@noop [0]{\@secondoftwo}%
\providecommand \href [0]{\begingroup \@sanitize@url \@href}%
\providecommand \@href[1]{\@@startlink{#1}\@@href}%
\providecommand \@@href[1]{\endgroup#1\@@endlink}%
\providecommand \@sanitize@url [0]{\catcode `\\12\catcode `\$12\catcode
  `\&12\catcode `\#12\catcode `\^12\catcode `\_12\catcode `\%12\relax}%
\providecommand \@@startlink[1]{}%
\providecommand \@@endlink[0]{}%
\providecommand \url  [0]{\begingroup\@sanitize@url \@url }%
\providecommand \@url [1]{\endgroup\@href {#1}{\urlprefix }}%
\providecommand \urlprefix  [0]{URL }%
\providecommand \Eprint [0]{\href }%
\@ifxundefined \urlstyle {%
  \providecommand \doi  [0]{\begingroup \@sanitize@url \@doi}%
  \providecommand \@doi [1]{\endgroup \@@startlink {\doibase
  #1}doi:\discretionary {}{}{}#1\@@endlink }%
}{%
  \providecommand \doi  [0]{doi:\discretionary{}{}{}\begingroup
  \urlstyle{rm}\Url }%
}%
\providecommand \doibase [0]{http://dx.doi.org/}%
\providecommand \Doi [0]{\begingroup \@sanitize@url \@Doi }%
\providecommand \@Doi  [1]{\endgroup\@@startlink{\doibase#1}\@@Doi}%
\providecommand \@@Doi [1]{#1\@@endlink}%
\providecommand \selectlanguage [0]{\@gobble}%
\providecommand \bibinfo  [0]{\@secondoftwo}%
\providecommand \bibfield  [0]{\@secondoftwo}%
\providecommand \translation [1]{[#1]}%
\providecommand \BibitemOpen [0]{}%
\providecommand \bibitemStop [0]{}%
\providecommand \bibitemNoStop [0]{.\EOS\space}%
\providecommand \EOS [0]{\spacefactor3000\relax}%
\providecommand \BibitemShut  [1]{\csname bibitem#1\endcsname}%
\bibitem [S1]{Mao2000.MaterResBull.35.1813.SM}%
  \BibitemOpen
  \bibfield  {author} {\bibinfo {author} {\bibfnamefont {Z.}~\bibnamefont
  {Mao}}, \bibinfo {author} {\bibfnamefont {Y.}~\bibnamefont {Maeno}},\ and\
  \bibinfo {author} {\bibfnamefont {H.}~\bibnamefont {Fukazawa}},\ }\Doi
  {10.1016/S0025-5408(00)00378-0} {\bibfield  {journal} {\bibinfo  {journal}
  {Mater. Res. Bull.}\ }\textbf {\bibinfo {volume} {35}},\ \bibinfo {pages}
  {1813} (\bibinfo {year} {2000})}\BibitemShut {NoStop}%
\bibitem [S2]{Mackenzie1998.PhysRevLett.80.161.SM}%
  \BibitemOpen
  \bibfield  {author} {\bibinfo {author} {\bibfnamefont {A.~P.}\ \bibnamefont
  {Mackenzie}}, \bibinfo {author} {\bibfnamefont {R.~K.~W.}\ \bibnamefont
  {Haselwimmer}}, \bibinfo {author} {\bibfnamefont {A.~W.}\ \bibnamefont
  {Tyler}}, \bibinfo {author} {\bibfnamefont {G.~G.}\ \bibnamefont
  {Lonzarich}}, \bibinfo {author} {\bibfnamefont {Y.}~\bibnamefont {Mori}},
  \bibinfo {author} {\bibfnamefont {S.}~\bibnamefont {Nishizaki}},\ and\
  \bibinfo {author} {\bibfnamefont {Y.}~\bibnamefont {Maeno}},\ }\Doi
  {10.1103/PhysRevLett.80.161} {\bibfield  {journal} {\bibinfo  {journal}
  {Phys. Rev. Lett.}\ }\textbf {\bibinfo {volume} {80}},\ \bibinfo {pages}
  {161} (\bibinfo {year} {1998})}\BibitemShut {NoStop}%
\bibitem [S3]{Deguchi2004RSI.SM}%
  \BibitemOpen
  \bibfield  {author} {\bibinfo {author} {\bibfnamefont {K.}~\bibnamefont
  {Deguchi}}, \bibinfo {author} {\bibfnamefont {T.}~\bibnamefont {Ishiguro}},\
  and\ \bibinfo {author} {\bibfnamefont {Y.}~\bibnamefont {Maeno}},\
  }\href@noop {} {\bibfield  {journal} {\bibinfo  {journal} {Rev. Sci.
  Instrum.}\ }\textbf {\bibinfo {volume} {75}},\ \bibinfo {pages} {1188}
  (\bibinfo {year} {2004}{\natexlab{a}})}\BibitemShut {NoStop}%
\bibitem [S4]{Deguchi2002.SM}%
  \BibitemOpen
  \bibfield  {author} {\bibinfo {author} {\bibfnamefont {K.}~\bibnamefont
  {Deguchi}}, \bibinfo {author} {\bibfnamefont {M.~A.}\ \bibnamefont
  {Tanatar}}, \bibinfo {author} {\bibfnamefont {Z.}~\bibnamefont {Mao}},
  \bibinfo {author} {\bibfnamefont {T.}~\bibnamefont {Ishiguro}},\ and\
  \bibinfo {author} {\bibfnamefont {Y.}~\bibnamefont {Maeno}},\ }\href@noop {}
  {\bibfield  {journal} {\bibinfo  {journal} {J. Phys. Soc. Jpn.}\ }\textbf
  {\bibinfo {volume} {71}},\ \bibinfo {pages} {2839} (\bibinfo {year}
  {2002})}\BibitemShut {NoStop}%
\bibitem [S5]{Deguchi2004.PhysRevLett.92.047002.SM}%
  \BibitemOpen
  \bibfield  {author} {\bibinfo {author} {\bibfnamefont {K.}~\bibnamefont
  {Deguchi}}, \bibinfo {author} {\bibfnamefont {Z.~Q.}\ \bibnamefont {Mao}},
  \bibinfo {author} {\bibfnamefont {H.}~\bibnamefont {Yaguchi}},\ and\ \bibinfo
  {author} {\bibfnamefont {Y.}~\bibnamefont {Maeno}},\ }\Doi
  {10.1103/PhysRevLett.92.047002} {\bibfield  {journal} {\bibinfo  {journal}
  {Phys. Rev. Lett.}\ }\textbf {\bibinfo {volume} {92}},\ \bibinfo {pages}
  {047002} (\bibinfo {year} {2004}{\natexlab{b}})}\BibitemShut {NoStop}%
\bibitem [S6]{Deguchi2004.JPhysSocJpn.73.1313.SM}%
  \BibitemOpen
  \bibfield  {author} {\bibinfo {author} {\bibfnamefont {K.}~\bibnamefont
  {Deguchi}}, \bibinfo {author} {\bibfnamefont {Z.~Q.}\ \bibnamefont {Mao}},\
  and\ \bibinfo {author} {\bibfnamefont {Y.}~\bibnamefont {Maeno}},\ }\Doi
  {10.1143/JPSJ.73.1313} {\bibfield  {journal} {\bibinfo  {journal} {J. Phys.
  Soc. Jpn.}\ }\textbf {\bibinfo {volume} {73}},\ \bibinfo {pages} {1313}
  (\bibinfo {year} {2004}{\natexlab{c}})}\BibitemShut {NoStop}%
\bibitem [S7]{Silhanek2006.PhysRevLett.96.136403.SM}%
  \BibitemOpen
  \bibfield  {author} {\bibinfo {author} {\bibfnamefont {A.~V.}\ \bibnamefont
  {Silhanek}}, \bibinfo {author} {\bibfnamefont {M.}~\bibnamefont {Jaime}},
  \bibinfo {author} {\bibfnamefont {N.}~\bibnamefont {Harrison}}, \bibinfo
  {author} {\bibfnamefont {V.~R.}\ \bibnamefont {Fanelli}}, \bibinfo {author}
  {\bibfnamefont {C.~D.}\ \bibnamefont {Batista}}, \bibinfo {author}
  {\bibfnamefont {H.}~\bibnamefont {Amitsuka}}, \bibinfo {author}
  {\bibfnamefont {S.}~\bibnamefont {Nakatsuji}}, \bibinfo {author}
  {\bibfnamefont {L.}~\bibnamefont {Balicas}}, \bibinfo {author} {\bibfnamefont
  {K.~H.}\ \bibnamefont {Kim}}, \bibinfo {author} {\bibfnamefont
  {Z.}~\bibnamefont {Fisk}}, \bibinfo {author} {\bibfnamefont {J.~L.}\
  \bibnamefont {Sarrao}}, \bibinfo {author} {\bibfnamefont {L.}~\bibnamefont
  {Civale}},\ and\ \bibinfo {author} {\bibfnamefont {J.~A.}\ \bibnamefont
  {Mydosh}},\ }\Doi {10.1103/PhysRevLett.96.136403} {\bibfield  {journal}
  {\bibinfo  {journal} {Phys. Rev. Lett.}\ }\textbf {\bibinfo {volume} {96}},\
  \bibinfo {pages} {136403} (\bibinfo {year} {2006})}\BibitemShut {NoStop}%
\bibitem [S8]{Lortz2007.PhysRevB.75.094503.SM}%
  \BibitemOpen
  \bibfield  {author} {\bibinfo {author} {\bibfnamefont {R.}~\bibnamefont
  {Lortz}}, \bibinfo {author} {\bibfnamefont {N.}~\bibnamefont {Musolino}},
  \bibinfo {author} {\bibfnamefont {Y.}~\bibnamefont {Wang}}, \bibinfo {author}
  {\bibfnamefont {A.}~\bibnamefont {Junod}},\ and\ \bibinfo {author}
  {\bibfnamefont {N.}~\bibnamefont {Toyota}},\ }\Doi
  {10.1103/PhysRevB.75.094503} {\bibfield  {journal} {\bibinfo  {journal}
  {Phys. Rev. B}\ }\textbf {\bibinfo {volume} {75}},\ \bibinfo {pages} {094503}
  (\bibinfo {year} {2007})}\BibitemShut {NoStop}%
\bibitem [S9]{NishiZaki2000JPhysSocJpn.SM}%
  \BibitemOpen
  \bibfield  {author} {\bibinfo {author} {\bibfnamefont {S.}~\bibnamefont
  {NishiZaki}}, \bibinfo {author} {\bibfnamefont {Y.}~\bibnamefont {Maeno}},\
  and\ \bibinfo {author} {\bibfnamefont {Z.}~\bibnamefont {Mao}},\ }\Doi
  {10.1143/JPSJ.69.572} {\bibfield  {journal} {\bibinfo  {journal} {J. Phys.
  Soc. Jpn.}\ }\textbf {\bibinfo {volume} {69}},\ \bibinfo {pages} {572}
  (\bibinfo {year} {2000})}\BibitemShut {NoStop}%
\bibitem [S10]{Tenya2006.JPhysSocJpn.75.023702.SM}%
  \BibitemOpen
  \bibfield  {author} {\bibinfo {author} {\bibfnamefont {K.}~\bibnamefont
  {Tenya}}, \bibinfo {author} {\bibfnamefont {S.}~\bibnamefont {Yasuda}},
  \bibinfo {author} {\bibfnamefont {M.}~\bibnamefont {Yokoyama}}, \bibinfo
  {author} {\bibfnamefont {H.}~\bibnamefont {Amitsuka}}, \bibinfo {author}
  {\bibfnamefont {K.}~\bibnamefont {Deguchi}},\ and\ \bibinfo {author}
  {\bibfnamefont {Y.}~\bibnamefont {Maeno}},\ }\Doi {0.1143/JPSJ.75.023702}
  {\bibfield  {journal} {\bibinfo  {journal} {J. Phys. Soc. Jpn.}\ }\textbf
  {\bibinfo {volume} {75}},\ \bibinfo {pages} {023702} (\bibinfo {year}
  {2006})}\BibitemShut {NoStop}%
\end{thebibliography}
\end{document}